# Super-resolution nanolithography of 2D materials by anisotropic etching

[1,2]Dorte R. Danielsen, [2,3]Anton Lyksborg-Andersen, [1,2]Kirstine E. S. Nielsen, [4]Bjarke S. Jessen, [1,2]Timothy J. Booth, [1,2]Peter Bøggild, and [1,2]Lene Gammelgaard

[1]Department of Physics, Technical University of Denmark (DTU), Kgs. Lyngby, Denmark.

[2]Centre for Nanostructured Graphene (CNG), Technical University of Denmark, Ørsteds Plads 345C, DK-2800 Kgs. Lyngby, Denmark

[3]DTU Nanolab - National Centre for Nano Fabrication and Characterization, Technical University of Denmark (DTU), Kgs. Lyngby, Denmark

[4]Department of Physics, Columbia University, New York, New York 10027, United States

E-mail: pbog@dtu.dk

Phone: +45 21 36 27 98

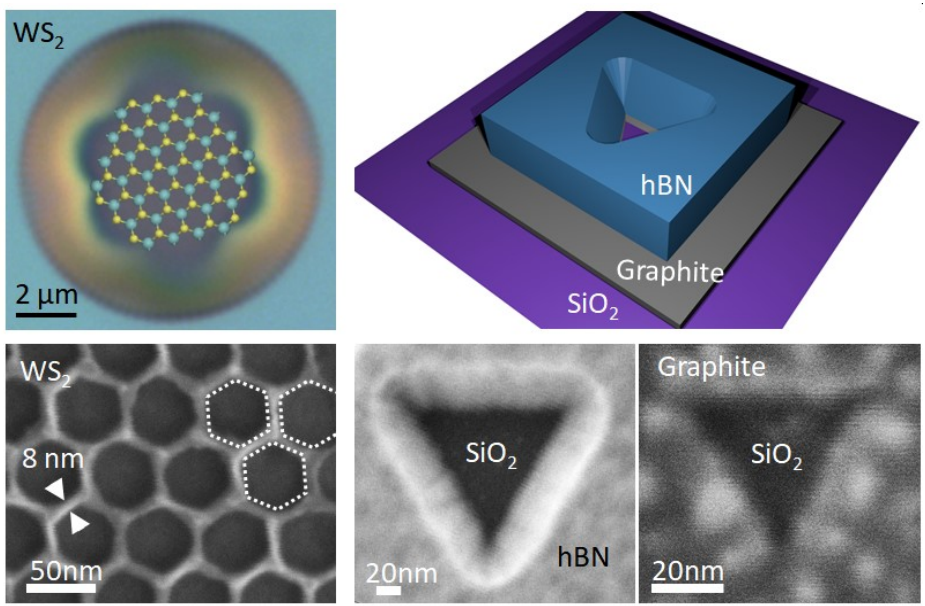

**Abstract**

Nanostructuring allows altering of the electronic and photonic properties of two-dimensional (2D) materials. The efficiency, flexibility, and convenience of top-down lithography processes are however compromised by nm-scale edge roughness and resolution variability issues, which especially affects the performance of 2D materials. Here we study how dry anisotropic etching of multilayer 2D materials with sulfur hexafluoride ($SF_6$) may overcome some of these issues, showing results for hexagonal boron nitride (hBN), tungsten disulfide ($WS_2$), tungsten diselenide ($WSe_2$), molybdenum disulfide ($MoS_2$), molybdenum ditelluride ($MoTe_2$). Scanning and transmission electron microscopy reveal that etching leads to anisotropic hexagonal features in the studied transition metal dichalcogenides, with the relative degree of anisotropy ranked as: $WS_2 >$ $WSe_2 > MoTe_2 \sim MoS_2$. Etched holes are terminated by zigzag edges while etched dots (protrusions) are terminated by armchair edges. This can be explained by Wulff constructions, taking the relative stabilities of the edges and the AA' stacking order into account. Patterns in $WS_2$ are transferred to an underlying graphite layer, demonstrating a possible use for creating sub-10 nm features. In contrast, multilayer hBN exhibits no lateral anisotropy, but shows consistent vertical etch angles, independent of crystal orientation. This is used to create super-resolution lithographic patterns with ultra-sharp corners at the base of the hBN crystal, which are transferred into an underlying graphite crystal. We find that the anisotropic $SF_6$ reactive ion etching process makes it possible to downsize nanostructures to obtain smooth edges, sharp corners, and feature sizes significantly below the resolution limit of electron beam lithography. The nanostructured 2D materials can be used themselves or as etch-masks to pattern other nanomaterials.

# 1 Introduction

Top-down nanostructuring of 2D materials is a fast and highly flexible route for achieving nanomaterials with specific electronic and optical properties. This is useful in multiple areas, such as plasmonics[1], bandstructure engineering[2, 3], quantum confinement[4], spin- and valley-tronics[5], sensing [6], and catalysis with transition metal dichalcogenides (TMDs)[7-9].

2D materials are, however, exceptionally sensitive to edge disorder and contamination, easily arising for standard protocols of nanopatterning as was pointed out early[10, 11] and later examined theoretically for antidot lattices[12]. A range of techniques have been explored to resolve this issue, including self-assembled resists[13, 14], scanning probe lithography[15, 16], direct beam lithographies[17, 18] as well as numerous resist-based lithographies. While bottom-up synthesis using molecular precursors may yield exceptional small feature sizes and low level of defects on the atomic scale, these materials are limited in terms of long-range order and freedom of design, as these structures rely on highly sophisticated and specific chemical synthesis routes[19-22]. Top-down electron beam lithography (EBL) provides long-range order and superior opportunities for pattern design, but is in practice limited to features sizes around 10 nm by secondary electron scattering effects, while edge roughness with optimization can approach 1 nm root mean square (RMS) roughness[23]. EBL combined with anisotropic etching, allows straightforward creation of well-defined structures with feature sizes significantly smaller than the EBL resolution limit, and with atomically smooth edges[24, 25]. This hybrid lithography approach is an enticing yet robust opportunity to achieve near-atomic-scale critical dimensions without compromising the freedom of design, inherent in top-down approaches.

Anisotropic etching of mono- and few-layer graphene as well as hBN, using a hydrogen plasma in a Chemical Vapor Deposition (CVD) system is well-studied[26-33]. Similarly, oxidative etching of $WSe_2$ and $MoS_2$ has been used to achieve anisotropically etched structures[7, 26, 34-37]. With this approach, the etch time is typically long (from multiple minutes to hours) and as the etching typically starts at random lattice defects, the nanostructures are not well-ordered. To resolve these issues, lithographic patterning with EBL and reactive ion etching (RIE) combined with subsequent $XeF_2$ etching has been used to create hexagonal holes in $MoS_2$[38]. Recently, wet etching with aqueous solutions of $H_2O_2$ and $NaH_4OH$ was used to create hexagonal patterns in $WS_2$, $MoS_2$ and molybdenum selenide ($MoSe_2$)[39] with near-atomic precision.

Here, we explore an alternative anisotropic RIE route for 2D materials, which eliminates the need for an extra etching step and ensures that etching only takes place at the E-beam defined spots. We distinguish between the in-plane (lateral) etching of the 2D layers (lateral isotropy/anisotropy), and the out-of-plane etching (vertical isotropy/anisotropy).

Using a pure sulfur hexafluoride ($SF_6$) plasma[2, 40], we etch several TMDs ($MoS_2$, $WS_2$, $WSe_2$, and $MoTe_2$), and find hexagonal hole geometries and nanoscale features locked to the crystalline directions similar to the recently published wet-etching results[39], however with large variations in the lateral anisotropy for the four studied TMDs.

In contrast, using the same etching process with hBN yield entirely different results. For hBN the dry $SF_6$ etch is laterally isotropic but with a very consistent etch angle confirming our previous observations[2]. We show here that this allows for a much wider palette of hole geometries and orientations compared to in-plane crystallographic etching[39], yet with similarly nm-sharp features, using a technique we here termed downsizing. The structures are examined by scanning electron microscopy (SEM) and transmission electron microscopy (TEM), and analyzed in terms of the relative stability of the 2D crystal edges. This provides a new method for realizing advanced device architectures with extremely fine features, using a conventional, convenient, deterministic fabrication process. The results are summarized in Figure 1.

| | hBN | WS$_2$ | WSe$_2$ | MoS$_2$ | MoTe$_2$ |
|---|---|---|---|---|---|
| **Vertical etch rate, [nm/min]** | 570 | 570 | 250 | 230 | 170 |
| **Vertical anisotropy, Etch angle [°]** | 66 | 83(ZZ) | 82(ZZ) | 85(ZZ) | 88(ZZ)/85(AC) |
| **Lateral anisotropy, $\gamma$ = rate$_{AC}$/rate$_{ZZ}$** | <1.02 | 1.31 | 1.23 | 1.04 | 1.03 |

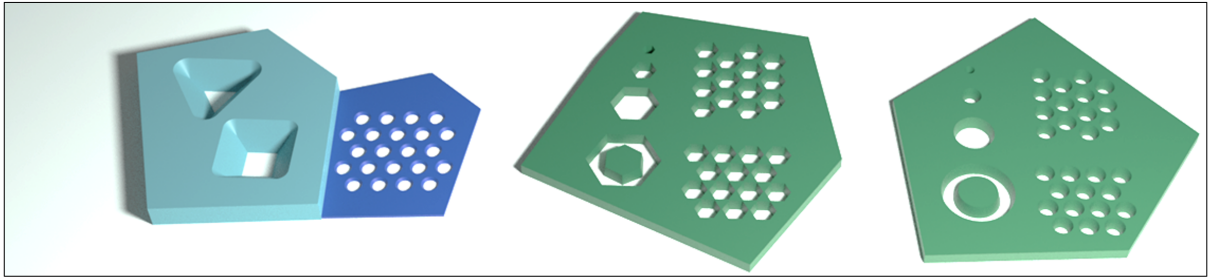

| | | | |
|---|---|---|---|
| **Shapes** | Any shape | Hexagonal holes/dots, nanowires | Resolution-limited |
| **Orientation** | Any orientation | Crystallographic | Any orientation |
| **Feature size** | Super-resolution (≤5 nm) | Super-resolution (≤ 5 nm) | Resolution-limited (>10 nm) |
| **Density** | Moderate | High | High |

Figure 1. Sub-resolution nanostructuring of 2D materials. The vertical etch rate, vertical etch anisotropy (etch angle) for ZZ and AC edges, and lateral anisotropy defined as the ratio $\gamma$ of etch rate $rate_{AC}$ in the AC direction to the etch rate to $rate_{ZZ}$ in ZZ direction, $\gamma = rate_{AC} / rate_{ZZ}$, for the five studied multilayer 2D crystals.

## 2  Dry etching and characterization

TMD and hBN crystals from hqgraphene.com, mechanically exfoliated onto 90 nm $SiO_2$, were used in this study. Crystals were manipulated and assembled into stacks using the hot pick-up dry transfer method[40]. The exfoliated flakes were patterned in a 100 keV JEOL JBX-9500FS E-beam system with a beam current of 0.8 nA. A 1.3 wt% solution of 2200 K Poly(methyl methacrylate) (PMMA) from Micro Resist Technology GmbH was used as a resist. Thin resist layers of 60 nm were found to provide an optimal compromise of resolution and etch resistance. A thicker resist layer of 140 nm (2wt% solution) was used for graphite-$WS_2$ and graphite-hBN heterostructures. The resist was developed in isopropanol (IPA) for 30 s, at room temperature. For further details on the fabrication process see Supplementary Section 1.

The crystals were etched in an inductively coupled plasma (ICP) system from SPTS Technologies (with pressure: 10 mTorr, $SF_6$ flowrate: 40 sccm, coil/platen power: 0/30 W, and time: 15 s - 90 s). Prior to etching, the samples were descummed for 5-15 s with a $O_2$ plasma to remove residual resist (pressure: 80 mTorr, $O_2$/Ar flowrates: 5/15 sccm, coil/platen power: 0/20 W). Graphite was etched with a $O_2$ plasma (pressure: 80 mTorr, $O_2$ flow rate: 20 sccm, coil/platen power: 0/20W, and time: 300 s). The $SF_6$ etch is highly selective and fully terminates on graphite[2, 40] allowing even monolayer graphene to act as an etch stop for atomically precise control of etching depths. Likewise, the pure $O_2$ etch is highly selective towards graphene and does not etch hBN or TMDs, allowing these to be used as complimentary hard masks for etching graphite and other nanomaterials or bulk surfaces. Details on etch rates and angles can be found in Supplementary Section 2.

The samples were characterized with SEM to visualize the etched patterns and determine the width of the sidewalls, using the in-lens detector and 5 kV acceleration voltage in a Zeiss Supra VP 40 SEM. To determine the vertical etch rates, the crystal thickness and step height of etched holes were measured with atomic force microscopy (AFM) using a Dimension Icon-PT from Bruker AXS in tapping mode. To perform TEM on the TMD crystals, the TMD crystals were transferred from the $SiO_2$/Si substrates to TEM grids, using cellulose acetate butyrate (CAB) wedging transfer with parameters similar to those in Ref. [41]. Both Cu TEM grids with amorphous carbon membrane and SiN membrane windows (SiMpore, 20nm SiN) were used. TEM imaging was performed with a Titan E-Cell 80-300ST TEM at 300 kV.

# 3 Results

## 3.1 Lateral etch rate and isotropy

To assess the lateral etching rate and isotropy, wagon wheel (WW) structures were etched in the TMDs and the hBN (Figure 2), following the approach in Ref. [42]. A WW structure, consists of triangular wedges radiating out from the center. During etching, the wedges are protected by the etch mask, but due to lateral etching under the resist, the width, $w$, is etched away from both sides of the wedges (Figure 2g). The wedges are designed to be sharp and narrow with a designed angle of $\varphi = 4°$. The small change in the width will therefore lead to the notable shortening of the wedge, $R$, compared to the design. $R$ and $w$ are related by $w = R \sin(\Delta\varphi / 2)$, and $R$ can be measured from SEM micrographs. The etch rate as a function crystal orientation angle, $r(\theta)$, can then be estimated as

$$r(\theta) = \frac{w(\theta)}{\Delta t} = \frac{R(\theta - 90°)\sin(\Delta\theta / 2)}{\Delta t}, \tag{1}$$

where $\Delta t$ is the etch time[42]. Note that the etch rate at a given angle, $\theta$, is determined by shortening of the wedges, $R$, orthogonal to $\theta$, as illustrated in Figure 2g.

Figure 2a-d i. show optical images of WWs with a radius of 5 µm in $WS_2$, $WSe_2$, $MoTe_2$, and $MoS_2$, respectively. Similarly, Figure 2a-d ii. show SEM micrographs of WWs with a radius of 2.5 µm. All structures are etched with $SF_6$ for 30 s. Green (long) and orange (short) arrows outline the fast and slow etching directions, respectively. Figure 2a-d iii. shows TEM images of holes that evolve into hexagons upon etching (left), together with the corresponding Fourier transformed (FT) images (right). Again, the fast and slow etching directions are indicated with green and orange arrows. The FT images have the same symmetry as diffraction images and can be used to determine the crystal orientation. On the FT images, the inner six spots labeled with orange circles are associated with the family of (1100) crystal planes corresponding to the zigzag (ZZ) direction, from which can be inferred that the hexagonal holes are terminated by ZZ edges.

The etch rate vs. crystal orientation angle is shown in a polar plot, see Figure 2e. The etch rates are estimated from SEM micrographs of WWs, including the images in Figure 2a-d ii., and the final etch rates shown on the graph are the mean values from at least six WWs for each material. The "fast" and "slow" etching directions are separated by 30°. The angular modulation of the etch rate can be expressed as the ratio $\gamma = rate_{AC} / rate_{ZZ}$ of etch rate in AC to the etch rate in ZZ direction.

The strongest anisotropy is observed for WS$_2$ ($\gamma_{WS_2} \approx 1.31$) and WSe$_2$ ($\gamma_{WSe_2} \approx 1.23$), while both MoTe$_2$ ($\gamma_{MoTe_2} \approx 1.03$) and MoS$_2$ ($\gamma_{MoS_2} \approx 1.04$) show very weak anisotropy. These results are summarized in Figure 1.

The TEM images in Figure 2a-d iii. show that the fast etching directions are orthogonal to the armchair (AC) edges. Correspondingly, the slow etching directions are orthogonal to the ZZ edges (see Figure 2f). The SEM and TEM images are not recorded from the same crystals, yet the correspondence between fast and slow etch rates and AC and ZZ edges is corroborated by TEM analysis of etch rates and crystal orientation in Supplementary Section 3.

Conversely, a large WW (5 µm radius) in hBN is shown in Figure 2h, exhibiting no indications of orientation dependent etch rate, confirming that the dry SF$_6$ etching of the hBN crystals is isotropic in the laterally direction[2].

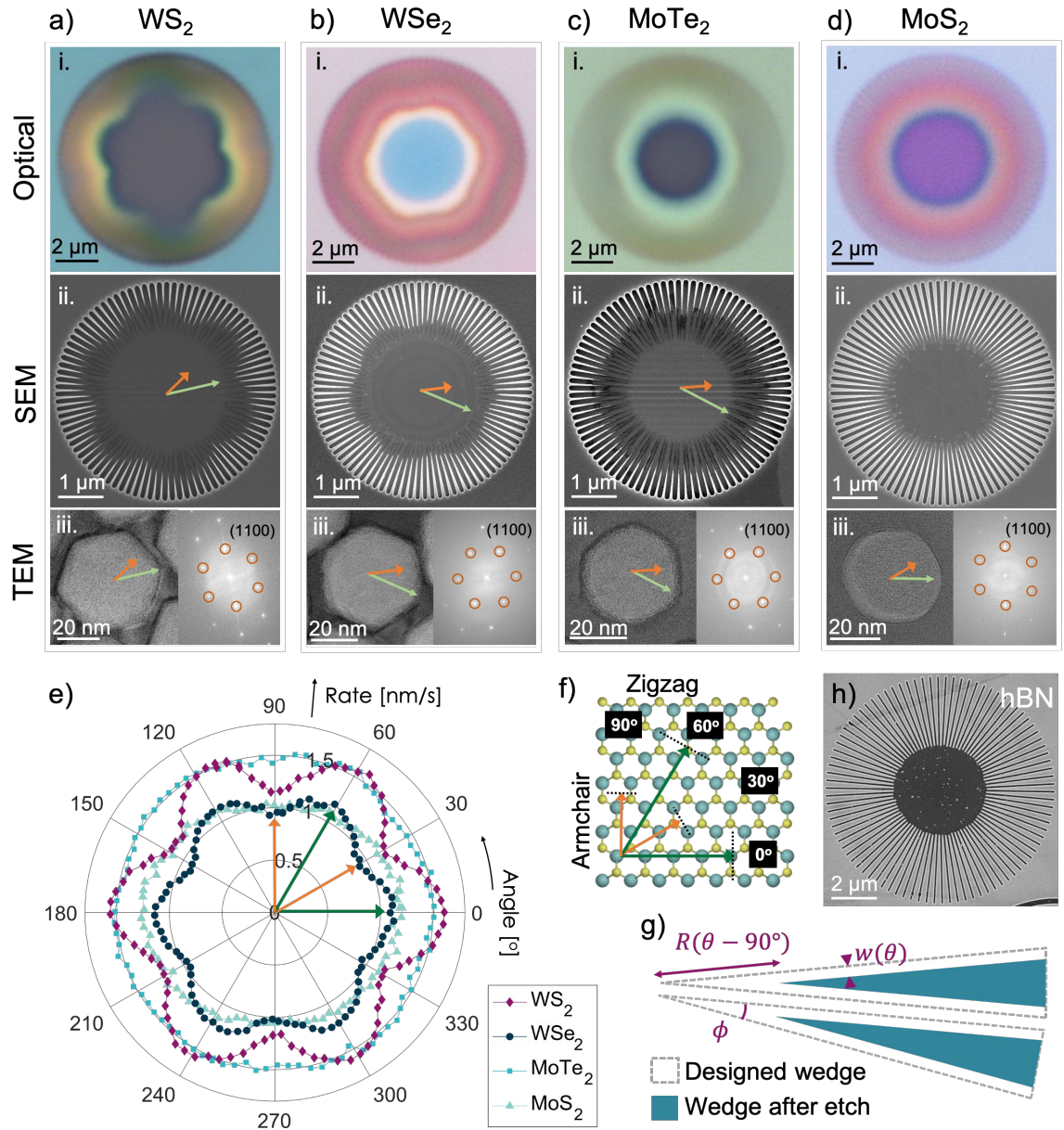

Figure 2. Anisotropic etching of TMDs. (a)-(d) i. Optical and (a)-(d) ii. SEM micrographs of WW structures in $WS_2$, $WSe_2$, $MoTe_2$, and $MoS_2$, respectively. The fast and slow etching directions are indicated by green (long) and orange (short) arrows, respectively. (a)-(d) iii. TEM images of hexagonal holes in the four TMD crystals, with FT of left images showing the same symmetry as the diffraction patterns of the TMD crystals. The inner spots labelled by orange circles belong to the family of (1100) spots, and the ZZ edges are orthogonal to the directions of these spots. (e) Polar plot of the etch rate as a function of crystal orientation angle, determined from SEM micrographs of the WWs. (f) Schematic illustration of the atoms in a TMD monolayer with the AC and ZZ directions indicated. The fast and slow etching directions are orthogonal to AC and ZZ edges, respectively. (g) Illustration of the wedges of the WW. (h) SEM micrograph of a WW in hBN.

## 3.2 Vertical etching rate and isotropy

The vertical etch rates and angles were measured using AFM, SEM, and TEM. With the etch process conditions described in section 2, the vertical etch rate for hBN was 570 nm/min, while the vertical etch rate for the TMDs was 170-250 nm/min, as listed in Figure 1. While the etch rate for hBN did not show any clear dependence on feature size, the etch rate for TMDs decreased for submicron features. For $WS_2$ the etch rate decreased from 200 nm/min for micron-scale features, to below 50 nm/min for nanoscale features, see Supplementary Figure 5.

For all structures, we determine the etch angle from measurements of the sidewall width with SEM and the height, which is measured by AFM (see Supplementary Section 2), and confirmed with TEM. Figure 3 shows TEM images of a "bull's-eye" structure and an antidot lattice etched in the same $WS_2$ crystal. The known thickness of the crystal and the TEM images of edge features (Figure 3b and c) allows accurate calculation of the etch angles. For $WS_2$, the etch angle differs for ZZ and AC edges. The etch angle of ZZ edges in the inner perimeter of the ring is 82.5°±1.0°, while the AC etch angle measured at the edges of the inner dot is 78.9°±1.5°. The vertical etching angle for hBN was measured to be 66°, in agreement with previous findings[2].

## 3.3 Nanopatterning with $WS_2$ and hBN

In the following, we discuss how the results presented in section 3.1 translates into nanopatterns of multilayer 2D crystals and subsequent pattern transfer to underlying substrates.

### 3.3.1 *Multilayer $WS_2$*

For the "bull's-eye" structure in Figure 3 the hexagonal inner and outer perimeters of the rings are rotated 30° with respect to each other, as seen in Figure 3a; this is a direct consequence of the crystallographically dependent etch rates, which we account for in section 4. Figure 3b and c show zoom-ins of the corners outlined by the blue rectangles, while Figure 3d shows the FT image of panel b. The orientation of inner spots in Figure 3d (labeled with orange circles) is orthogonal to the ZZ crystal directions and the outer spots (green circles) are orthogonal to the AC orientation. From this, it is evident that the inner perimeters of the bull's-eye rings are terminated by ZZ edges and the outer perimeters are terminated by AC edges. Figure 3e shows a regular hexagonal antidot lattice and Figure 3f is a zoom-in of the region outlined by the white box in Figure 3e. Since the bull's-eye structure and antidot lattice are etched in the same crystal, Figure 3d also represents the crystal symmetry of the antidot lattice. Thus, the hexagonal holes in the antidot lattice are ZZ terminated.

The corners in Figures 3b and c have radii of curvature on the order of 15 nm, with the corner for the smaller AC terminated dot in the center being sharper than the inner and outer corners of the larger ring, and small structures of the antidot lattice in Figures 3f have even smaller radius of curvature. The fact that smaller structures evolve into more sharply defined hexagons is also apparent in Supplementary Figures 22, 23, and 24.

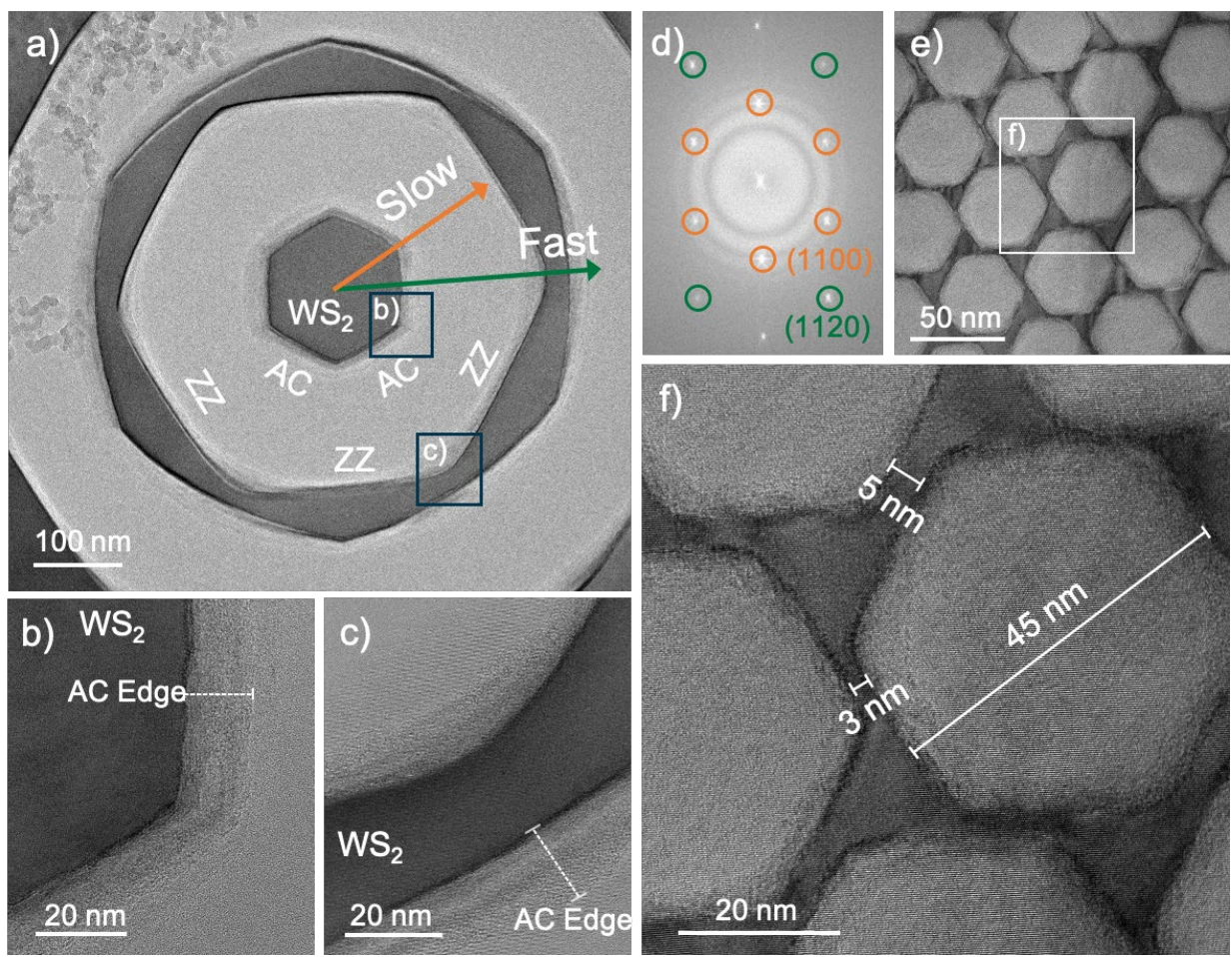

Figure 3. TEM images of bull's-eye structures and antidot lattices in WS$_2$. (a) Bull's-eye structure in WS$_2$ with the fast and slow etching directions indicated by long (green) and short (orange) arrows, respectively. (b)-(c) Zoom-in of the AC and ZZ corners outlined by the blue boxes in panel (a), allowing the width of the AC edge and the edge roughness to be estimated. (d) FT of the TEM image in panel (b). The inner spots (orange circles) can be associated with the family of (1100) planes, i.e. the ZZ crystal orientations, and the outer spots (green circles) can be associated with the family of (1120) planes, i.e. AC orientations. This means that the inner perimeters of the bull's-eye structure are ZZ-terminated and the outer perimeters are AC-terminated (see white labels in (a)). (e) Hexagonal antidot lattice with very regular features. (f) Zoom-in of the area outlined by the white box in (e). The holes are not etched completely through the crystal, which is clear from the crystalline atomically resolved pattern in the holes. The width of the hexagons is 45 nm and the constrictions between holes are ca. 3 nm.

The roughness of the AC edge in Figure 3b has contributions from the curvature of the edge, which appears to follow a parabolic envelope compared to the "true" AC edge, superposed with random roughness variations (see Supplementary Section 4 ). For the $WS_2$ AC edge, the maximum deviation from the AC edge facet is 5 nm while the root mean square (RMS) edge roughness is 0.36 nm, after the envelope has been subtracted (see Supplementary Figure 14). The roughness of a ZZ edge in the antidot lattice in Figure 3f is determined in a similar way. Here the maximum deviation of the edge from the true ZZ facet is 1.2 nm and the RMS roughness is 0.225 nm, when subtracting the envelope (see Supplementary Figure 15). The width of the hexagons in this lattice is 45 nm while the spacing between holes is as small as 3 nm. The holes are not etched completely through, which is apparent from the crystalline atomic pattern within. Due to the orientation of the antidot pattern with respect to the crystal lattice, a network of triangular islands connected by narrow nanoribbons is formed. If the antidot and crystal lattices are fully aligned or anti-aligned, a network of triangles or a honeycomb-like network of nanoribbons is formed, respectively (see Supplementary Figure 19), similar to Munkhbat et al. [39] for anisotropic wet etching of $WS_2$.

### 3.3.2 *Multilayer hBN*

The outcome of pattern downsizing a polygon with dry etching of hBN relies on trivial geometrical relations between the radius of corner curvature $r$ resulting from the finite resolution of the lithography process, the vertical angle $\phi$, the thickness $t$ of the hBN crystal, and the designed $D$ and resulting $d$ side lengths of the polygon, which is depicted for a triangle in Figure 4a. The minimum thickness, $h < t$, to achieve optimal sharpness of the corners is related to $r$ and $\phi$ as $h = r \tan \phi$. For our lithography process we get without any corner optimization or proximity correction, $r \approx 30$ nm , with which the RIE $SF_6$ constant vertical etch angle $\phi = 66°$, results in approximately $h = 70$ nm . Figure 4c and d show the results of etching acute ( $30°$ ) corners in two different hBN crystals, where the 31 nm crystal is clearly too thin, while the 78 nm results in excellent corner sharpness. Figure 4e shows the differences in shape and size for triangles, squares and hexagons etched in hBN with thicknesses larger than $h$. The polygons for both $t = 78$ nm and $t = 89$ nm all have sharp corners, but different sizes. For the $t = 120$ nm crystal, the holes are not etched through.

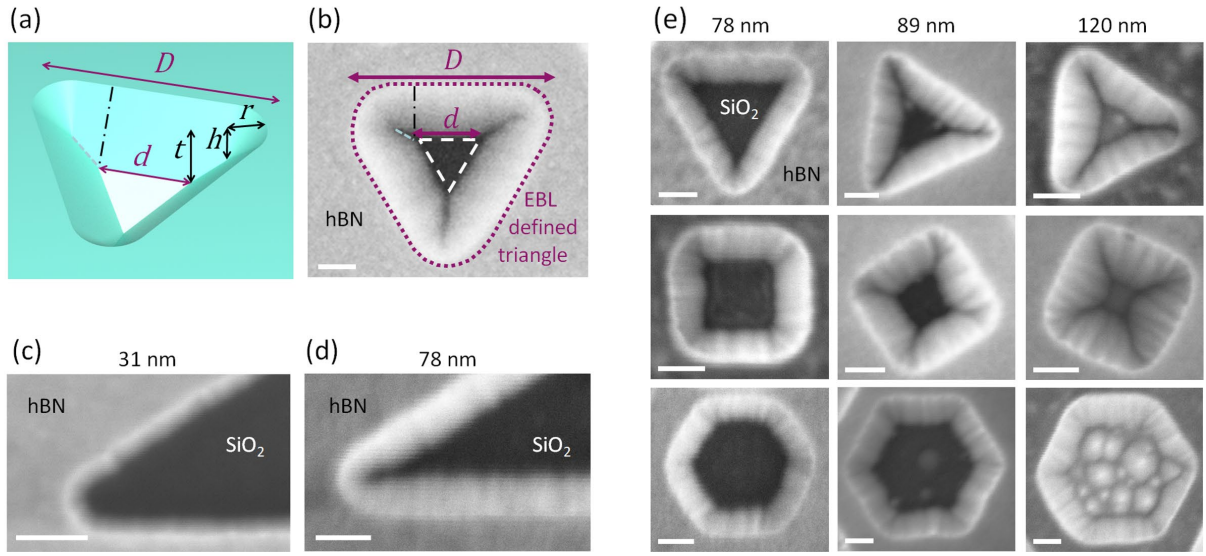

Figure 4. Geometric downsizing of triangles. (a) Schematic illustration of a triangular etched hole in hBN and the interrelationship between the geometric parameters. (b) SEM micrograph of a triangular hole in hBN. Due to the consistent etch angles (isotropic etch angles) in hBN, patterns at the base of the hBN crystal are downsized and have ultra-sharp corners. (c, d) An acute (30°) corner etched into a 31 nm and a 78 nm thick hBN crystal, where only the latter is thick enough to give a sharp corner. (e) SEM images of downsized polygons in hBN crystals with different thicknesses (78 nm, 89 nm, and 120 nm). The 120 nm crystal is not etched fully through, which is seen from the uneven bottom of the structures. All scale bars are 50 nm.

### 3.4 Pattern transfer with hBN and WS$_2$ hard masks

Pattern transfer and downsizing are visualized in both stacks of graphite-hBN and graphite-WS$_2$, as shown in Figure 5. Various structures are defined by EBL in a PMMA mask and then etched with the SF$_6$ RIE process into the hBN/WS$_2$, and thereafter into the underlying graphite with an oxygen RIE process, using the hBN/WS$_2$ as a hard masks. The crystallographically independent etch angle, $\phi$, and the lateral isotropic etch rate in hBN are used to downsize structures of arbitrary shapes and achieve sharp corners at the base of the hBN crystal in a deterministic way, i.e. getting highly predictable outcomes. The downsizing and sharpening effect requires that the angle of the corner is below 180° and the hBN crystal is thicker than the depth, $h$, where two sidewalls of a corner meet. The lateral anisotropic etching of WS$_2$ is used to transfer both hexagonal holes and dots into graphite.

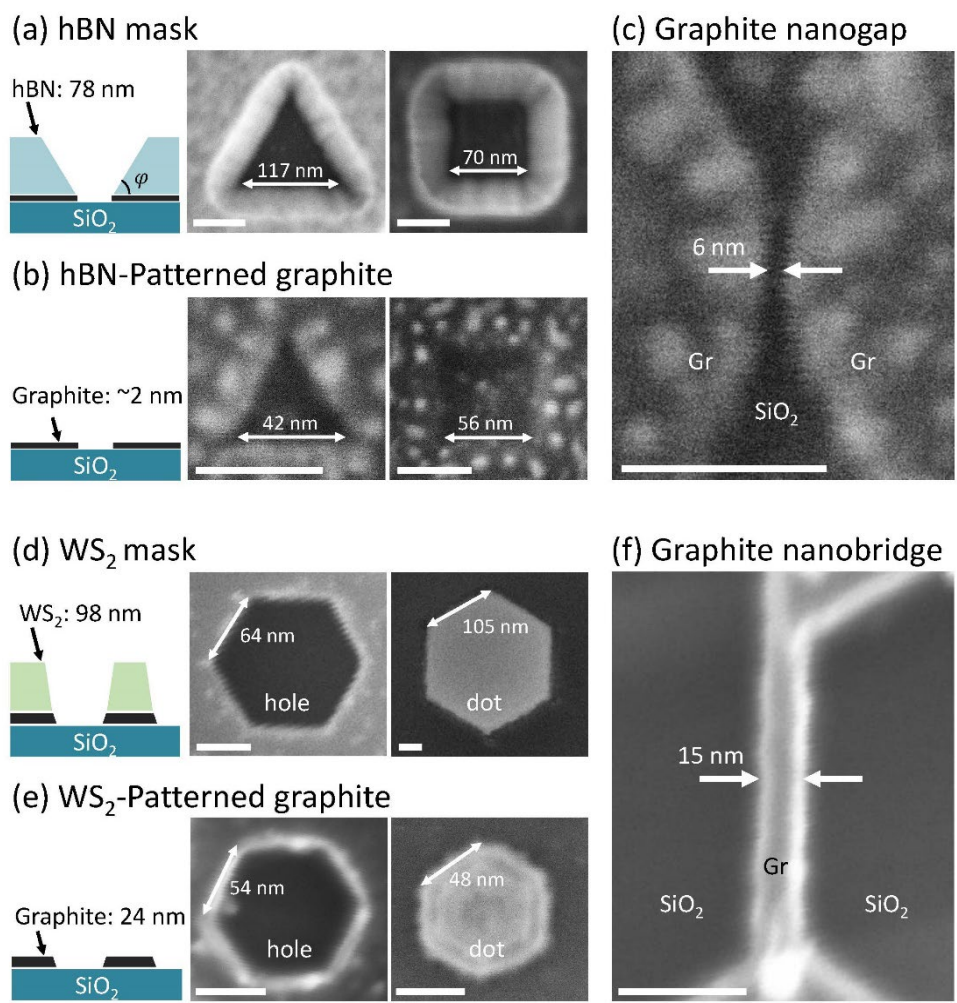

Figure 5. Pattern transfer. (a) Side-view schematic illustration of an etched hole in a graphite-hBN stack, with SEM micrographs of triangular and square holes etched in hBN. (b) Graphite holes after removal of hBN by further dry etching. (c) Example of 6 nm nanogap etched in graphite by a hBN hard mask. (d-e) Illustration and SEM micrographs of WS$_2$-patterning of hexagonal graphite holes and dots (d) before and after (e) removal of the WS2 mask. (f) Graphite nanobridge with a width of 15 nm defined by etching through WS2 hard masks consisting of two adjacent hexagonal holes. All scale bars are 50 nm.

Figure 5 shows schematics and SEM micrographs of the high-resolution pattern transfer process with hBN and WS$_2$. The graphite-hBN and a graphite-WS$_2$ stacks after etching the hBN/WS$_2$ with SF$_6$ are shown in Figure 5a and Figure 5d, respectively, with SEM images of triangular, square holes for hBN and hexagonal holes and dots for WS$_2$. The hard mask is then removed with the SF$_6$ RIE process, which is highly selective towards graphite[2, 40] leaving the structures etched into the graphite layer, as shown in Figure 5b (hBN) and Figure 5e (WS$_2$). Figure 5c shows the hBN down-sizing technique is used to create a 6 nm gap in two 2 nm thick graphite sheets, while Figure 5f shows a 15 nm wide bridge created in graphite using a WS$_2$ hard mask with two adjacent hexagonal holes. The width of the graphite bridge is measured from the SEM secondary electron line scan, using the distance between the peaks originating from the edge-bloom effect. The SEM micrographs taken before and after removing the hBN/WS$_2$ hard mask are from the same stack, but from different areas, as the SEM imaging leaves carbonaceous depositions, which prevents further processing[43]. See Supplementary Section 5 for more examples of transferred and downsized structures and Supplementary Figures 2 and 3 for an outline of the fabrication steps.

## 4 Discussion

We have examined anisotropic lateral and vertical dry etching in multilayer 2D materials. For multilayer TMDs, we note a striking difference in etch characteristics of the four studied MX$_2$ crystals compared to the previously published study of anisotropic wet etching[39].

From analysis of the WW structures in Figure 2, it is clear that WS$_2$ exhibits the largest degree of lateral anisotropy, followed by WSe$_2$, MoTe$_2$, and MoS$_2$. This observation is supported by EBL-defined circles and triangles which evolve into hexagons upon etching of the TMDs (see Supplementary Section 6). Even though we find hBN and MoS$_2$ to etch overall isotropically, a slight onset of hexagonal facets appears for very small etched features. This is seen for the 41 nm diameter hole in MoS$_2$ in Figure 2d.iii.

In the following we provide an intuitive explanation of the differences in lateral etch rates in terms of Wulff constructions[44, 45], which have previously been used to explain the etch dynamics of hydrogen etching of CVD-grown *monolayer* hBN and graphene[27, 28]. After etching, inner perimeters (holes and inner edges of bull's-eye rings) should be terminated by the slowest etching edges, while the outer perimeters (dots and outer edges of bull's-eye rings) should be terminated by the fastest etching edges[28] (see Figure 6a).

This argument can be extended to multilayer MX$_2$, by taking their stacking order into account. The most stable configuration for multilayer WSe$_2$ and other TMDs (MoS$_2$, MoSe$_2$, and WS$_2$) is AA' stacking, which is typical for the 2H phase[46]. According to density functional theory, the most stable edge for TMDs is the chalcogen-terminated ZZ edge[47, 48]. For monolayers, this implies that triangular holes with chalcogen-terminated ZZ edges should form, which is consistent with multiple studies of anisotropic etching of individual TMD layers[7, 35-37]. For multi-layers, the AA' stacking order of TMDs means that the preferentially etched chalcogen-terminated triangles will be rotated 180° in alternating layers[7, 36, 37], which then results in hexagonal ZZ-terminated holes in the TMD, by a mechanism depicted in Figure 6b. Similar explanations were given for multi-layer anisotropic etching of $MoS_2$ by selective steam-etching[7], and multi-layer anisotropic etching of $WS_2$ by chemical wet etching[39], while we here provide confirmation of this puzzling mechanism for a dry etching process. Similarly, the $SF_6$ etching process yields hexagonal ZZ-terminated holes and inner perimeters (see Figure 2a-d iii. and Figure 3) as well as AC-terminated dots and outer perimeters (see Figure 3a-d). The WWs in Figure 2 show that etching is fastest at the AC edge and slowest at the ZZ edge, so the results are consistent with the Wulff construction method[28].

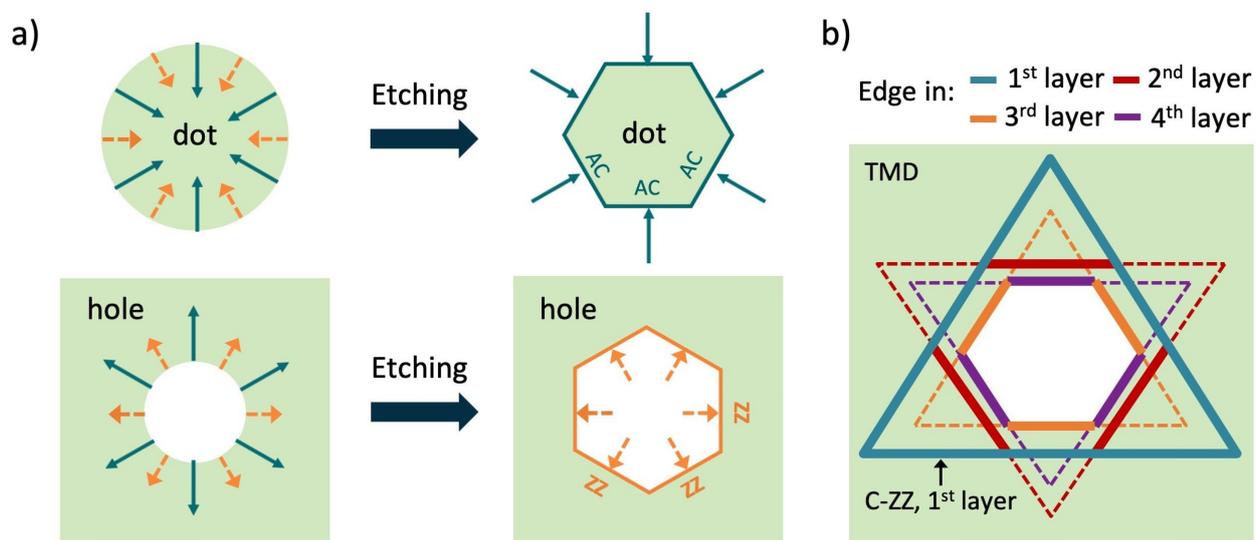

Figure 6. Edge terminations after anisotropic etching. (a) Following Wulff construction arguments made by Stehle et al.[28] etched holes will be terminated by the slowest etching edge (the most stable edge) and dots will be terminated by the fastest etching edge (the least stable edge). (b) For TMDs, the most stable edge appears to be the chalcogen-terminated ZZ edge, while the least stable edge is the AC edge. For multilayers with the common AA' stacking order, this leads to hexagonal holes and dots rotated 30° with respect to each other, in agreement with the argumentation in Ref. [7, 28].

The varying degree of etch anisotropy of $WS_2$, $WSe_2$, $MoTe_2$, and $MoS_2$ may arise from differences in the relative stabilities of AC and ZZ edges for the TMDs. This would fit with the Wulff construction picture[28] of the etching process. While it is generally true that the etch dynamics differ somewhat for different TMDs, it is tempting to suggest that tungsten plays a role in the anisotropic $SF_6$ etching, since $WS_2$ and $WSe_2$ exhibited the largest degree of anisotropic etching, compared to the Mo-TMDs; further studies are needed to test this hypothesis. An interesting feature of the anisotropic etching process is that the WW structures make it possible to determine the crystal directions of $WS_2$ and $WSe_2$ from optical images, which otherwise requires more elaborate techniques, such as TEM diffraction. This allows the crystal orientation to be determined while the crystal is already deposited on its target substrate, using just a single, non-critical EBL step, which does not in any way prevent further processing.

The RMS edge roughness of the structures in $WS_2$ is on the order of a few atoms (0.36 nm for the AC edge of the inner dot in Figure 3a, and 0.225 nm for the ZZ edge of the hexagonal antidot lattice in Figure 3f. The lower RMS edge roughness for the ZZ edge may result from the holes in the $WS_2$ antidot lattice not being etched all the way through. Our previous findings[2] showed that the presence of an underlying 2D layer tends to reduce edge roughness when etching hBN with the $SF_6$ RIE process, compared to etching through to a $SiO_2$ substrate. This trend may be exploited in future nanocircuitry and nanocomponents by incorporating 2D multilayer materials as etch stops to decrease not just vertical depth but also lateral feature integrity.

The low edge roughness achieved in this study is comparable to the edge roughness obtained by Munkhbat et al. [39] using RIE and subsequent anisotropic wet etching. While this wet etching process etches $MoS_2$ and $MoSe_2$ with somewhat stronger anisotropy, we find it striking that similar anisotropic etching and shaping can be achieved by a single, short RIE step. RIE processes ultimately rely on ion bombardment and are often more likely to give strong vertical rather than lateral anisotropy. We presume that the relatively low pressure and low platen power of our $SF_6$ etching process promotes a more "soft" chemical etching, allowing for the observed pronounced lateral anisotropic behaviour to emerge. Furthermore, the $SF_6$ plasma is highly selective with respect to PMMA (see Supplementary Figure 4), which is very important for creating dense patterns with small feature sizes.

In line with the different degrees of lateral anisotropic etching the etch profile also depend on the material. The etch angle in $WS_2$ are significantly different for AC and ZZ edges. A possible explanation is that the well-defined AC and ZZ edges in TMDs favor different edge facet geometries. This possibility is discussed further in Supplementary Section 2.3. $WS_2$ allows for dense arrays of nanobridges to be formed, either to be used as sub-10 nm nanowires or as etch masks for underlying layers or surfaces. These structures will always be locked to the crystal orientation and with shape limitations dictated by the in-plane crystallographic etching. In contrast, the etching behaviour of hBN differs from the studied TMDs in that hBN shows practically no signs of lateral anisotropy, and exhibits consistent vertical etch angles, independent of crystal orientation. As a result, the E-beam defined features appear downsized and with sharpened corners at the base of the etched hBN crystal. With proper matching of hBN thickness and lithography parameters, this enables hBN to be used as a tunable etch mask to easily define EBL patterns with sub-resolution features and arbitrary pattern geometries in an underlying layer, as demonstrated here with hBN on graphite.

In summary, we presented here a strategy to create super-resolution lithographic patterning of 2D materials using hBN as a self-sharpening hard-mask with dry $SF_6$ RIE etching. This approach can be extremely powerful, since it is well-known that hBN encapsulation provides phenomenal protection of the 2D materials even despite aggressive patterning[2, 4], hence, this approach uniquely provides superior protection and pattern resolution at the same time. We also demonstrated anisotropic dry etching of TMDs using RIE with $SF_6$, leading to very regular well-defined hexagonal structures with ultra-smooth edges and features sizes consistently below 10 nm, as obtained here for hexagonal anti-dot lattices, and demonstrated pattern transfer to graphite with similar dimensions.

The ability to downsize and sharpen features with hBN as an etch mask, and to use TMDs both as the active layer or as nanometer-scale etch masks pushes the limits for 2D nanostructuring. We anticipate that anisotropic dry etching provides a valuable toolkit for making advanced device architectures of high quality, with extreme density and near-atomic feature sizes, applicable in plasmonics, metamaterials, spin- and valleytronics, bandstructure engineering, and catalysis.

**Acknowledgement**

The authors thank Nicolas Stenger, Søren Raza, Antti-Pekka Jauho, Mads Brandbyge for helpful discussions, and Niels Pichon, Emil Duegaard and Joachim Søderquist for early assistance with experiments.

**Supporting Information Available**

Fabrication details, etch angle and etch rate analysis, TEM analysis of wagon wheels, edge roughness analysis, additional examples of pattern transfer, additional TEM, and SEM images.

**Super-resolution nanolithography of 2D materials by anisotropic etching – Supplementary Information**

[1,2]Dorte R. Danielsen, [2,3]Anton Lyksborg-Andersen, [1,2]Kirstine E. S. Nielsen, [4]Bjarke S. Jessen, [1,2]Timothy J. Booth, [1,2]Peter Bøggild, and [1,2]Lene Gammelgaard

[1]Department of Physics, Technical University of Denmark (DTU), Kgs. Lyngby, Denmark.

[2]Centre for Nanostructured Graphene (CNG), Technical University of Denmark, Ørsteds Plads 345C, DK-2800 Kgs. Lyngby, Denmark

[3]DTU Nanolab - National Centre for Nano Fabrication and Characterization, Technical University of Denmark (DTU), Kgs. Lyngby, Denmark

[4]Department of Physics, Columbia University, New York, New York 10027, United States

E-mail: pbog@dtu.dk

Phone: +45 21 36 27 98

# 1 Fabrication details

**Preparation of 2D crystals.** TMD crystals were mechanically exfoliated onto 90 nm $SiO_2$ grown on standard Si wafer chips, using scotch tape (3M Scotch ®810 Magic™), while hBN was exfoliated using cleanroom blue tape (EPL BT-150E-KL as mold-tape and SWT 20+ LB B 140.00 MM as copy-tape, both by Nitto Denko Corporation). Prior to exfoliation, the $SiO_2$/Si chips were heated on a hotplate for five minutes at 200°C, to remove water, and cleaned in oxygen plasma for five minutes (at approximately 300mbar $O_2$ and 120W in the Plasma Etch PE50 system). Immediately after plasma treatment, tape covered with 2D material flakes was brought in contact with the $SiO_2$ surface, and the round end of a pen was used to raster scan over the tape surface to improve adhesion. The tape was then slowly released from the surface (at room temperature for scotch tape and at 85°C for the SWT 20+ LB B 140.00 MM tape), while keeping a low angle between the $SiO_2$ surface and the tape.

**Electron beam lithography.** The exfoliated flakes were patterned with EBL, using poly(methyl methacrylate) (PMMA) as a resist. PMMA (1.3 wt% of 2200K PMMA from Micro Resist

Technology GmbH) was spin coated on the chip (1500 rpm, 500 rpm/s, for 60 s with a 10min post-bake at 180 °C), forming a 60 nm thick layer. A thicker resist layer of 140 nm (2wt% solution spun at 1500 rpm, 500 rpm/s, for 60 s) was used for the pattern transfer samples where longer etch times are needed. The E-beam exposure was performed with a 100 keV JEOL JBX-9500FS system, with a beam current of 0.8nA. The hexagonal anti-dot lattices were defined with single shot exposures using total charge doses ranging from $1.5 \times 10^{-15} - 9.1 \times 10^{-15}$ C/dot while the other structures were defined with regular (areal) exposures using doses of 2000-5000 µC/cm2. The resist was developed in IPA for 30 s, at room temperature.

**Sample types.** Patterns are etched in two sample types: 1) Exfoliated crystals of TMDs and hBN, 2) stacks of either hBN or $WS_2$ transferred on top of graphite. Wagon wheels, bull's eyes, antidot lattices, and nanostructures of various shapes are defined by E-beam lithography in a PMMA mask and etched into the 2D materials. The process of nanostructuring exfoliated crystals is shown in Supplementary Figure 1. The exfoliated TMDs and hBN are etched with the $SF_6$ RIE process. SEM inspection is performed after the final fabrication step in Supplementary Figure 1d.

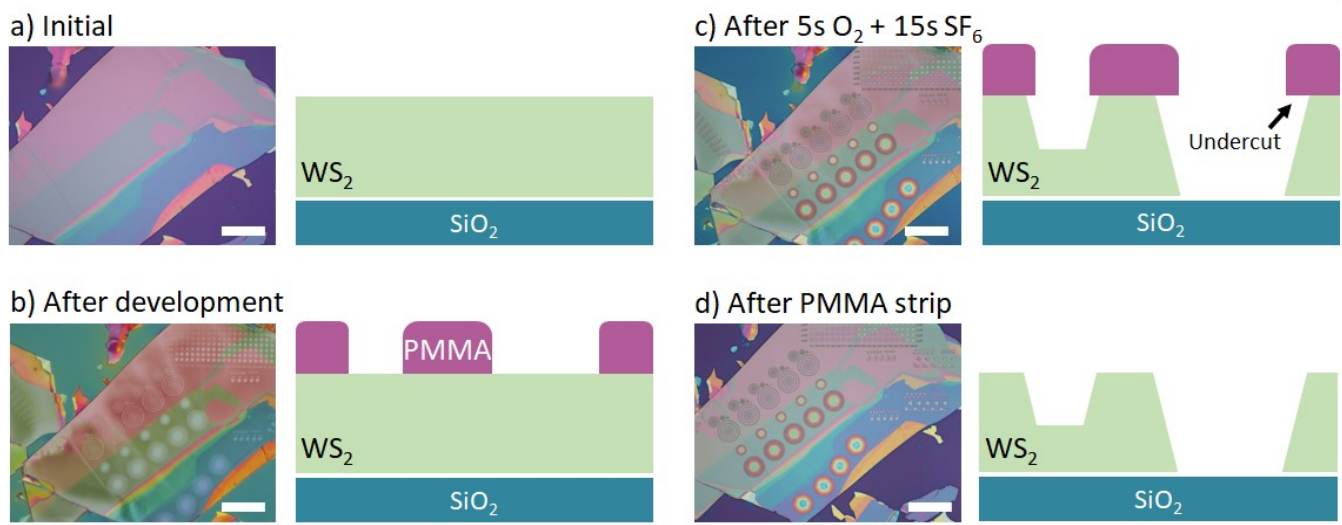

Figure 1. Nanopatterning process. Optical micrographs of a $WS_2$ crystal during nanopatterning along with schematic illustrations the shape of the etch profile is unknown and are here depicted straight. (a) Initial $WS_2$ crystal. (b) Crystal after patterning by EBL and development. 60nm 2200k PMMA, a current of 0.8 nA, and 30s of development in IPA is used. (c) After etching the $WS_2$, with a 5s $O_2$ PMMA descum followed by a 15s $SF_6$ RIE process. Both the vertical etch rate, and the lateral undercut depend on the dimension of the etched structures, crystal thickness and the etch time. (d) Nano-structured crystal after removing the PMMA in acetone. Large structures in this $WS_2$ crystal are not etched fully to the base of the crystal in the relative short etch time. Scale bars are 20µm.

Fabrication of samples for demonstration of the downsizing and sharpening with a pattern transfer through hBN and WS$_2$ are presented in Supplementary Figure 2 and Figure 3, respectively. The hBN/WS$_2$ is etched with SF$_6$ and graphite is etched with the O$_2$ RIE process.

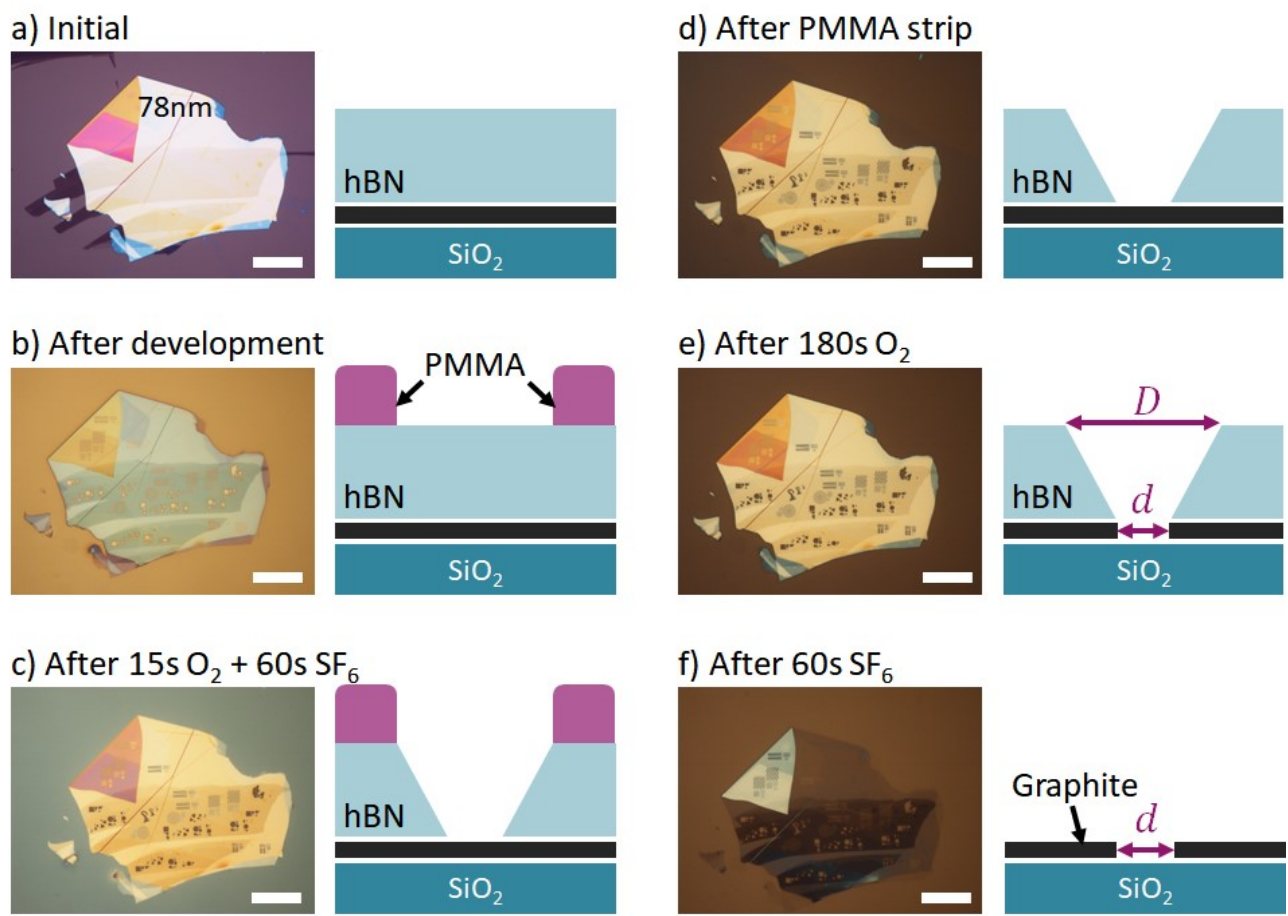

Figure 2. Pattern transfer process with a hBN mask. Optical micrographs along with schematic illustrations of a graphite-hBN stack during the fabrication steps for demonstrating downsizing and pattern sharpening. (a) Graphite-hBN stack after stacking using the hot pick-up method[1]. (b) Stack after patterning by EBL and development. 140nm 2200k PMMA, a current of 0.8 nA, and 30s of development in IPA is used. (c) After etching the hBN, with a 15s O$_2$ PMMA descum followed by a 60s SF$_6$ RIE process. (d) Nanostructured stack after removing the PMMA in acetone. (e) After etching the graphite with O$_2$ in 180s. (f) The hBN is removed with a 60s SF$_6$ etch, a part of the hBN where the crystal is folded on top of itself is not etched away, the rest is fully etched. A color difference is observed on the SiO$_2$, as some of the oxide is etched during the 180s O$_2$ etch. Scale bars are 20μm.

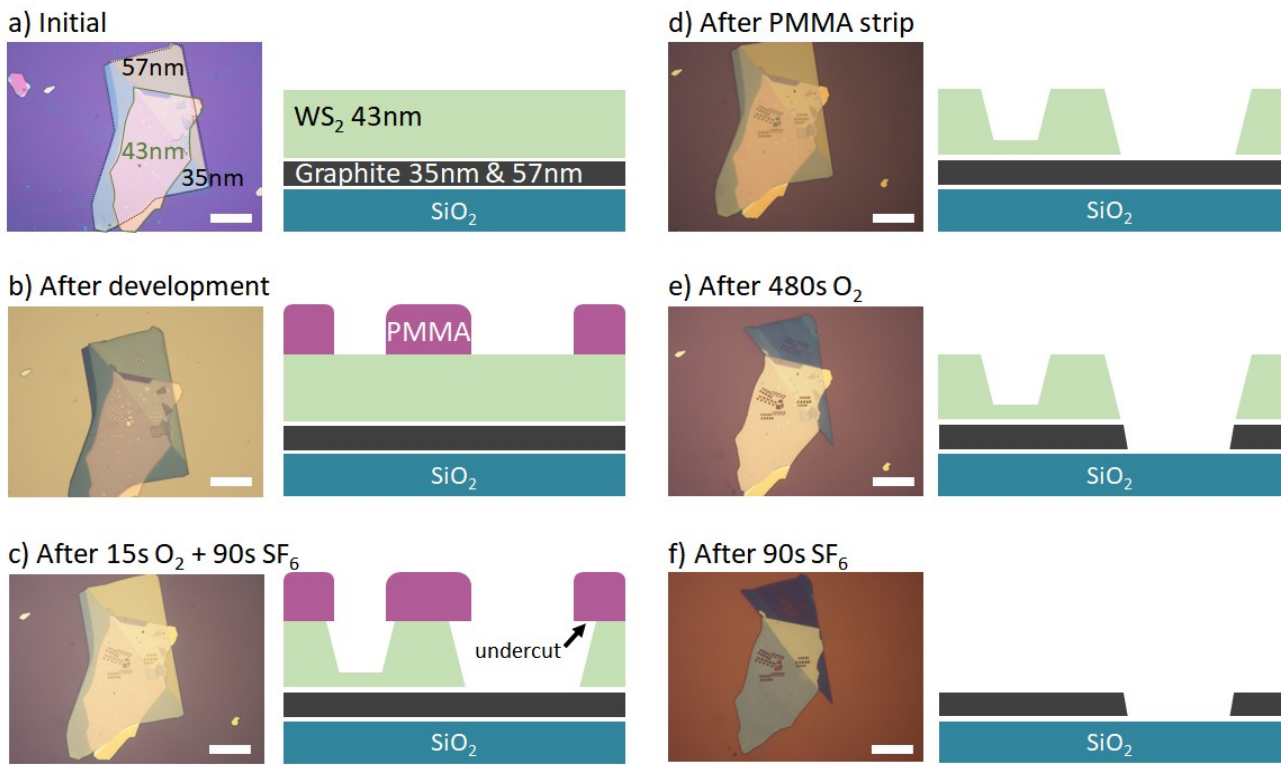

Figure 3. Pattern transfer process with a $WS_2$ mask. (a) Graphite-$WS_2$ stack after stacking using the hot pick-up method[1]. The $WS_2$ is 43 nm thick and the graphite have two regions, one that is 35 nm thick and one that is 57 nm thick. (b) Stack after patterning by EBL and development. 140 nm 2200k PMMA, a current of 0.8nA, and 30s of development in IPA is used. (c) After etching the $WS_2$, with a 15s $O_2$ PMMA descum followed by a 90s $SF_6$ RIE process. (d) Nanostructured stack after removing the PMMA in acetone. (e) After etching the relative thick graphite with $O_2$ in 480s. (f) The $WS_2$ is removed with a 90s $SF_6$ etch. Scale bars are 20μm.

## 2 Determining Etch Angles and Etch Rate

### 2.1 Etch rate dependence on feature size and crystal thickness.

The vertical etch rates in various materials for the $SF_6$ RIE process used throughout this study are listed in Supplementary Figure 4. The etch is selective towards TMDs and hBN and does not etch graphene or graphite. The etch in PMMA is low compared to the TMDs and hBN, which makes it possible to use a relatively thin PMMA layer, ensuring a good pattern transfer from the PMMA resist mask to the TMD/hBN.

| Material | WS$_2$ | WSe$_2$ | MoS$_2$ | MoTe$_2$ | hBN | Gr | PMMA |
|---|---|---|---|---|---|---|---|
| Vertical etch rate [nm/min] | 230 | 250 | 230 | 170 | 570 | 0 | 80 |

Figure 4. Vertical etch rates for the SF$_6$ RIE process. Etch rates of µm size areas in different materials. The etch is highly selective towards hBN and TMDs, has a low etch rate in the E-beam resist, PMMA, and terminates on graphene/graphite layers.

The etch rates listed in Supplementary Figure 4 are for µm size areas. The etch rate for small features can be significantly lower, and thicker PMMA is therefore used for the pattern transfer samples to enable longer etch times and pattern transfer of small structures. We have observed that the vertical etch rate for WS$_2$ depends strongly on the feature size as seen in Supplementary Figure 5. Similar trends are observed for the other TMDs, where small features are not etched fully to the bottom of the crystal when larger areas are (see Supplementary Figure 3 and Figure 20).

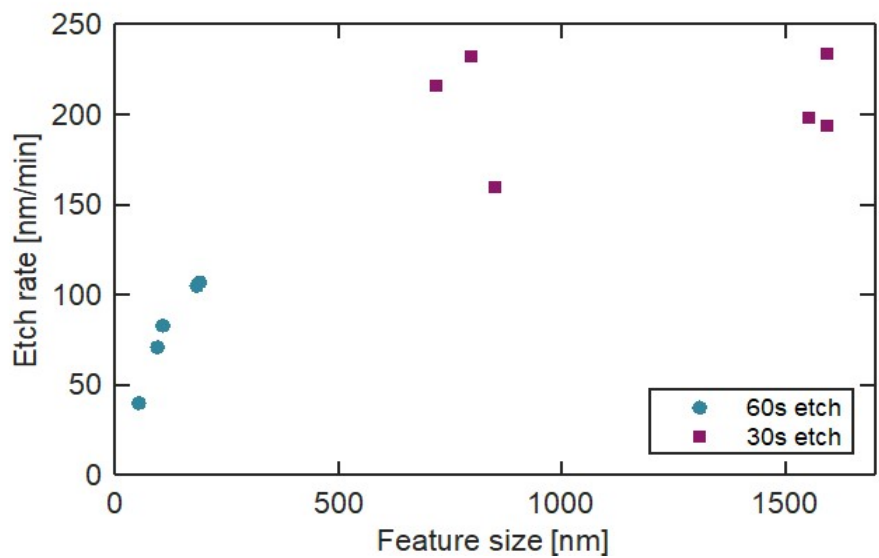

Figure 5. Vertical etch rate in WS$_2$ dependency on the feature size of the structure. The etch rate is lower for small structures and plateaus for large features. The step height of etched holes are measured by AFM on two different crystals, etched for 30s and 60s.

The etch rate in hBN does not seem to be notably affected by feature size, however, the fast vertical etch rate of 570 nm/min in hBN, and the etch angle of 66° lead to small features evolving quickly into a cone-like shape. We do not observe a lateral undercut for hBN, and small features will therefore terminate at the cone-like shape, even when increasing the etch time. In contrast, the anisotropic etching of the TMDs leads to a noticeable undercut. The undercut of ZZ-terminated edges versus the designed diameter of round holes is plotted in Supplementary Figure 6 for WS$_2$ crystals of different thicknesses and various etched times. The undercut is determined by measuring the side-to-side length in SEM micrographs of the etched hexagons and comparing this to the designed dimensions. The gradual etching of the PMMA mask over time is not taken into account. The data presented in Figure 6 are for samples etched over a time span of half a year, so run-to-run variations are expected and make it difficult to compare different runs. Generally, the undercut distance increases rapidly for small diameters and plateaus for large diameters. Longer etch times also lead to a larger undercut. Two flakes from the same chip but of different thicknesses (43 nm and 98 nm) are etched together for 90 s (blue circles and triangles). The lateral undercut of the thinner crystal (43 nm) terminates at approximately 54 nm, while the thicker crystal exhibits a larger undercut of 65 nm for the largest features. This suggests that the lateral underetching slows down significantly when the space under the PMMA is too confined.

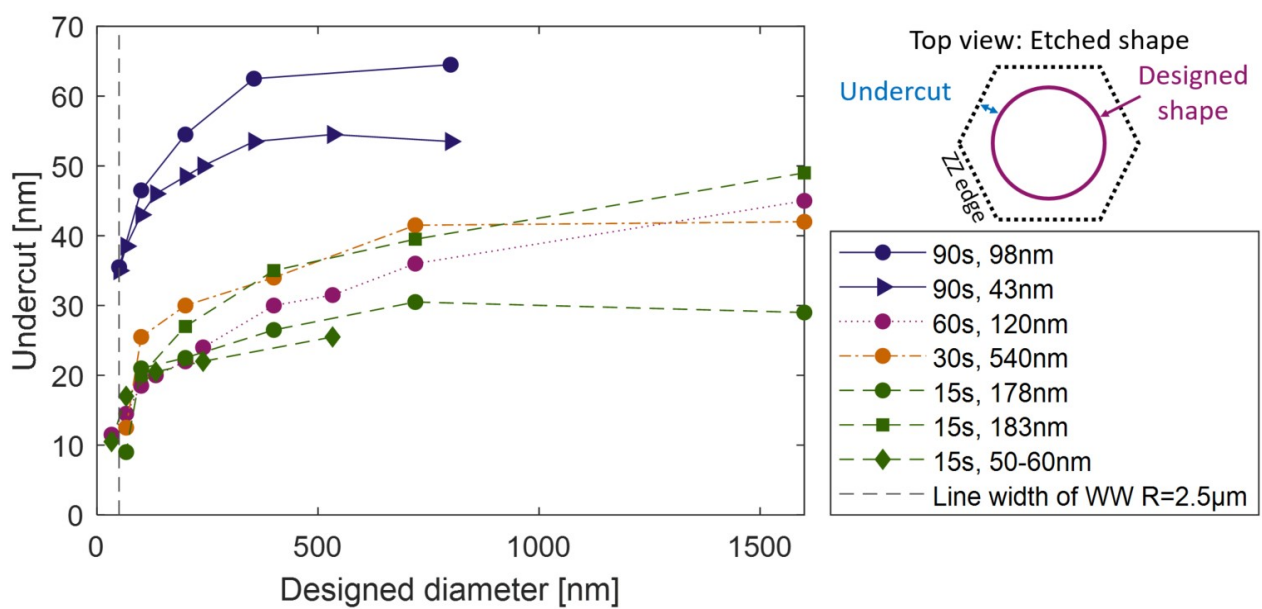

Figure 6. Lateral undercut. The lateral undercut at the ZZ terminated edge vs. the designed diameter of a circular hole. The undercut distance is measured from SEM micrographs of flakes of various thickness etched from 15 s to 90 s.

The interplay between the vertical and lateral etch rate, together with feature size, crystal thickness, and etch time all influence the final dimensions of an etched structure in a $WS_2$ crystal. These parameters should therefore be optimized for specific designs, and kept constant for reproducibility.

## 2.2 Vertical etch rate dependence on feature size

In many of the fabricated antidot lattices (Supplementary Figure 3 and Figure 20) the holes are not etched fully through the TMD crystal, even though larger structures in the same crystals are. It turns out that the vertical etch rate depends strongly on the feature size of the etched structure, with smaller patterns having a lower etch rate compared to larger structures (see Supplementary Figure 5). Feature-size dependent etching is a well-known phenomenon in RIE processes[2, 3] and is linked to the transport of ions and neutral species. Furthermore, the lateral anisotropic undercut responsible for circular holes evolving into hexagons depends on the feature size and the thickness of the crystal. The lateral undercut of the ZZ edge for various $WS_2$ crystals and etch times can be seen in Supplementary Figure 6. Generally, we find that smaller features have smaller lateral undercuts, and the undercut is smaller for thinner crystals than for thicker crystals, etched at the same time. This might be a result of the lateral etching slowing down when it obtains a hexagonal structure and the edges are terminating at the slower etching ZZ edges. Conversely, more material needs to be etched away to turn large structures into hexagons[4] leading to larger undercuts, but also not fully evolved hexagons. The maximum diameter of circles that evolve into hexagons, therefore, depends on the degree of lateral anisotropic etching, the etch time, and the crystal thickness. Here, large anisoropy, long etch times, and thick crystals lead to larger structures evolving into hexagons. For more information see Supplementary section 6.2.1.

## 2.3 Etch angle dependence on crystal orientation.

The etch angle, $\phi$, can be calculated from the step height, $h$, and the width of the edge, $w$, as

$$\varphi = \tan^{-1}\left(\frac{h}{w}\right). \qquad (2)$$

The step height in the TMDs and hBN is measured with AFM, while SEM is used to measure the edge width in hBN. The etch angle in the 135 nm thick hBN crystal in Supplementary Figure 7 is measured to be $66 \pm 1°$. This is in line with previous results for hBN etching with similar $SF_6$ etching parameters, where the etch angle was $66.1°\pm0.6°$, and independent of crystallographic orientation[5].

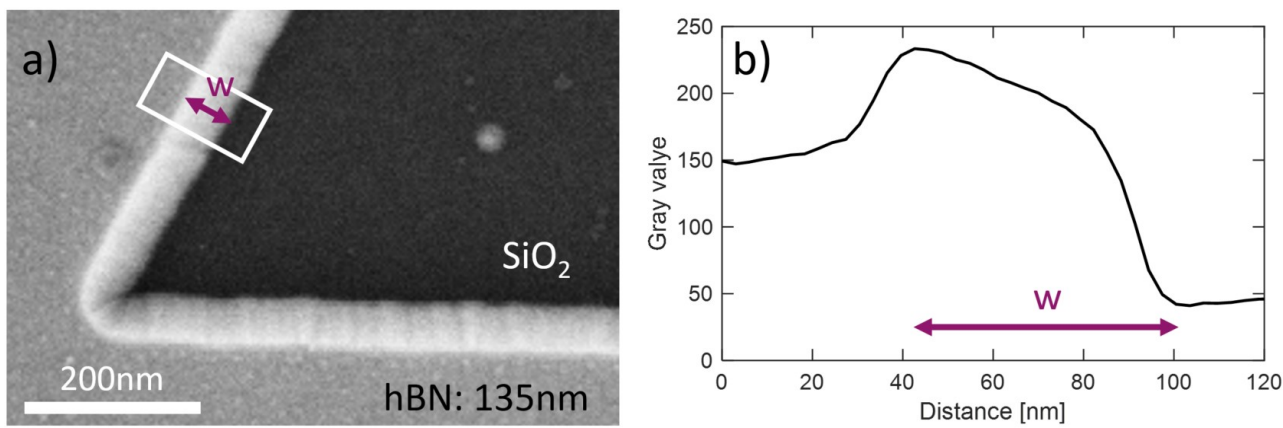

Figure 7. Etch angle in hBN. (a) SEM micrograph of an etched hBN crystal, the hBN is by AFM measured to be 135nm thick. (b) Gray-scale line scan in the box indicated in (a). The double ended arrow indicates the width of the edge. The etch angle is calculated to be 66 ± 1°.

Even though hBN generally shows no signs of lateral etching anisotropy and no etch angle dependency on the crystal orientation[5], we do observe initial signs of evolving facets for the small dots in Supplementary Figure 21c. The models in Supplementary Figure 8 illustrates how the AC and ZZ edge facets might look if hBN is etched anisotropically. In model one the AC edge facet contains pure AC edges in each layer (Supplementary Figure 8a). In contrast, the ZZ edge facet in model one only has nitrogen-terminated ZZ edges exposed at the edge facet (Supplementary Figure 8b). The reasoning behind this is that the nitrogen-terminated ZZ edge is the most stable edge [4, 6, 7]. In model two (Supplementary Figure 8c) the ZZ edge facet contains alternating boron- and nitrogen terminated ZZ edges. If the energy difference between the boron- and nitrogen terminated ZZ edges is insufficient in favoring nitrogen-terminated edges, this would be the expected edge facet geometry. For TMDs the models can also be used, with the nitrogen-terminated ZZ edges replaced by chalcogen-terminated ZZ edges.

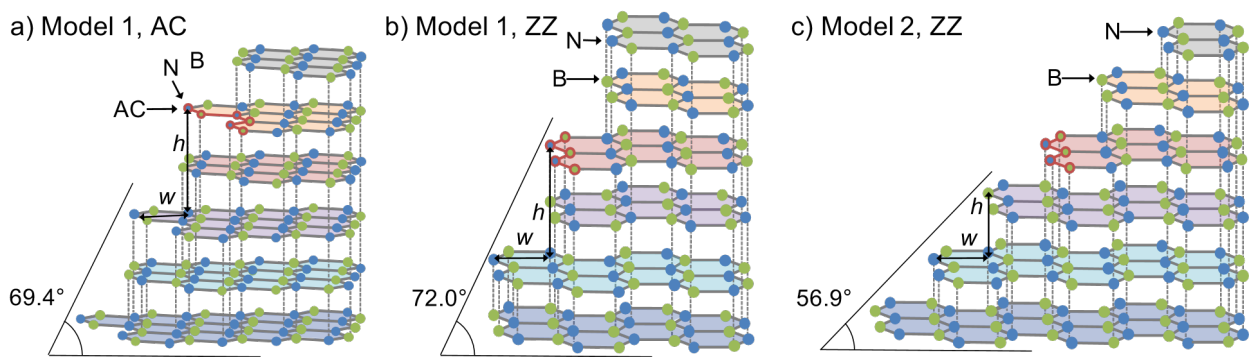

Figure 8. Etch angle models. (a)-(c) Illustrations of the edge facets in two proposed models. For each model, the etch angle is calculated from the lattice constants a and c, where a is measured from the FT TEM images, and c is from http://www.hqgraphene.com.

Based on these models, the proposed etch angles can be calculated by letting $h$ be the height of the unit cell containing two atomic layers (usually denoted $c$), and w be the width of one atomic step from two layers (see eq. (2)). w is determined from the width of the unit cell (usually denoted $a$). For AC edges, $w = a$, and for ZZ edges, $w = a \sin 60°$, as illustrated in Supplementary Figure 9.

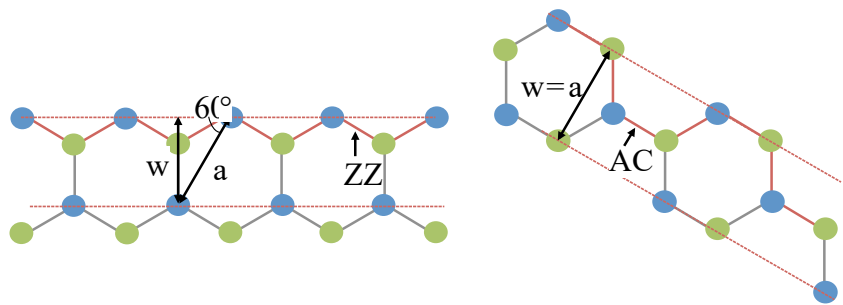

Figure 9. Sketch of the atomic step width. w is here the width of an edge consisting of two atomic layers. For ZZ edges, $w = a \sin 60°$, and for AC edges, $w = a$, where a is the lattice constant.

For hBN, the measured etch angle is close to the proposed angles but it does not match any of them. This makes sense since the lateral etching is mostly isotropic, so the edges are not expected to precisely follow the AC or ZZ edge facets.

In WS$_2$ we observed a significantly higher etch angle for the ZZ edges compared to the AZ edges both when inspecting crystals by SEM and TEM. The measured and model values of the etch angles are listed in Supplementary Figure 10.

| WS$_2$ | AC etch angle [°] | ZZ etch angle [°] | Thickness [nm] |
|---|---|---|---|
| Model 1 | 75.6 | 77.5 | - |
| Measured (TEM) | 78.9 ± 1.5 | 82.5 ± 1.0 | 74 |
| Measured (SEM) | 78 ± 2 | 84 ± 2 | 98 |

Figure 10. Etch angles of WS$_2$. Calculated and measured etch angles for the ZZ and AC edges in WS$_2$, along with the thickness of the crystal etched. The crystal inspected by TEM was etched for 30s with SF$_6$, and the crystal inspected by SEM was etched for 90s with SF$_6$. Both crystal were etched fully through in this process.

TEM is used to measure the etch angles of the bull's eye in Main Figure 3. The width of the edge region is indicated on Main Figure 3b-c. The AC etch angle is measured from the edges of the inner dot, while the ZZ etch angle is measured from the edges at the inner perimeter of the first ring. The 74nm crystal was etched for 30 s with $SF_6$. The etch angle has been measured in the SEM micrographs of the bull's eye in Supplementary Figure 11 by examining the change in the grayscale value over the edge of the structure. The bull's eye is etched into a stack of graphite-$WS_2$ only the $WS_2$ have been etched at the time of the SEM inspection. The $WS_2$ have been etched first with a 15 s $O_2$ PMMA descum followed by a 90s $SF_6$ RIE process. The obtained angle is the average of the six AC edges of the center dot in the bull's eye and the average of the six ZZ edges at the inner perimeter of the first ring of the structure. Examples of the change in the grayscale value is seen in Supplementary Figure 11c. The ZZ angle is difficult to measure from the SEM micrograph since the grayscale values change very abruptly, limited by the resolution of the SEM. The measured value for the ZZ etch angle of approximately 84° ± 2° should therefore be taken as the minimal value. The actual ZZ etch angle may be higher. Note that the inner dot in the bull's eye appears to have rotated slightly from the 30° misalignment with the inner perimeter of the ring. We expect this to be a result of the small $WS_2$ dot snapping to the lattice orientation of the underlying graphite layer.

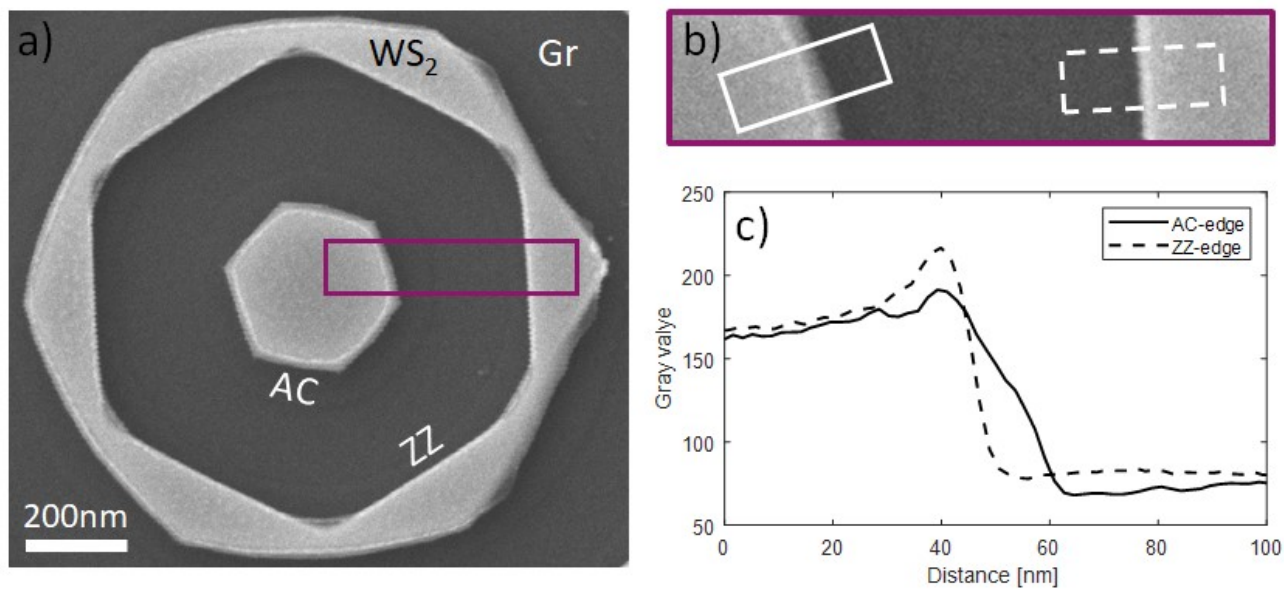

Figure 11. Etch angle in $WS_2$. (a) SEM micrograph of a bull's eye structure etched into a stack of 98 nm thick $WS_2$ on a 24 nm thick graphite crystal. Only the $WS_2$ is etched. (b) Zoom-in on the AC edge of the center dot and ZZ edge of the inner side of the first ring. (c) Average grayscale value in the white boxes indicated in (b). The average horizontal distance for the six AC edges of the dot is 21 nm and the six ZZ inner edges of the ring is 10 nm, leading to estimated etch angles of approximately 78° ± 2° for the AC edges and a minimum angle of 84° ± 2° for the ZZ edges.

For WS$_2$, the proposed etch angles obtained using Model 1 in Supplementary Figure 8 also deviate from the measured etch angles, however, the model capture the trend of higher etch angles for ZZ edges than AC edges. Model 1 predict etch angles of 75.6°/77.5° for the AC/ZZ edges, respectively. Previously, Wang et al. [8] obtained high-resolution AFM images of anisotropically etched holes in MoS$_2$, showing that the atomic steps along the sidewall always consist of an even number of atomic layers, except for the top and bottom layer. Wang et al. [8] often observed more than two layers in each step, which could also explain the higher measured etch angles in this study, compared to the angles in the proposed models where all steps are considered to be of two layers.

## 3 Crystal Orientation in Wagon Wheels

To confirm the correspondence between AC and ZZ edges, and fast and slow etching directions, the etch rate is calculated from TEM images of wagon wheels, where Fourier transformed TEM images can directly identify the crystal directions of the wagon wheels (see Supplementary Figure 12). For each TMD the etch rates in Supplementary Figure 13 are calculated from a single wagon wheel, so it is not an average. It is still evident that WS$_2$ and WSe$_2$ exhibit anisotropic etching with fast and slow etching directions separated by 30°. Based on the TEM images in Supplementary Figure 12, the exact crystal orientation at a given position angle on the image can be determined. This is used to rotate the individual wagon wheels such that the zero-angle on the graph identifies the direction orthogonal to the AC edge for all wagon wheels. For WS$_2$ and WSe$_2$ it is clear that the etch rate peaks and 0°, 60°, 120°, etc. and is lowest at 30°, 90°, 150°, etc. This confirms that the fastest etching direction is the direction orthogonal to AC edges (long green arrow), and the slowest etching direction is the direction orthogonal to ZZ edges (short orange arrows).

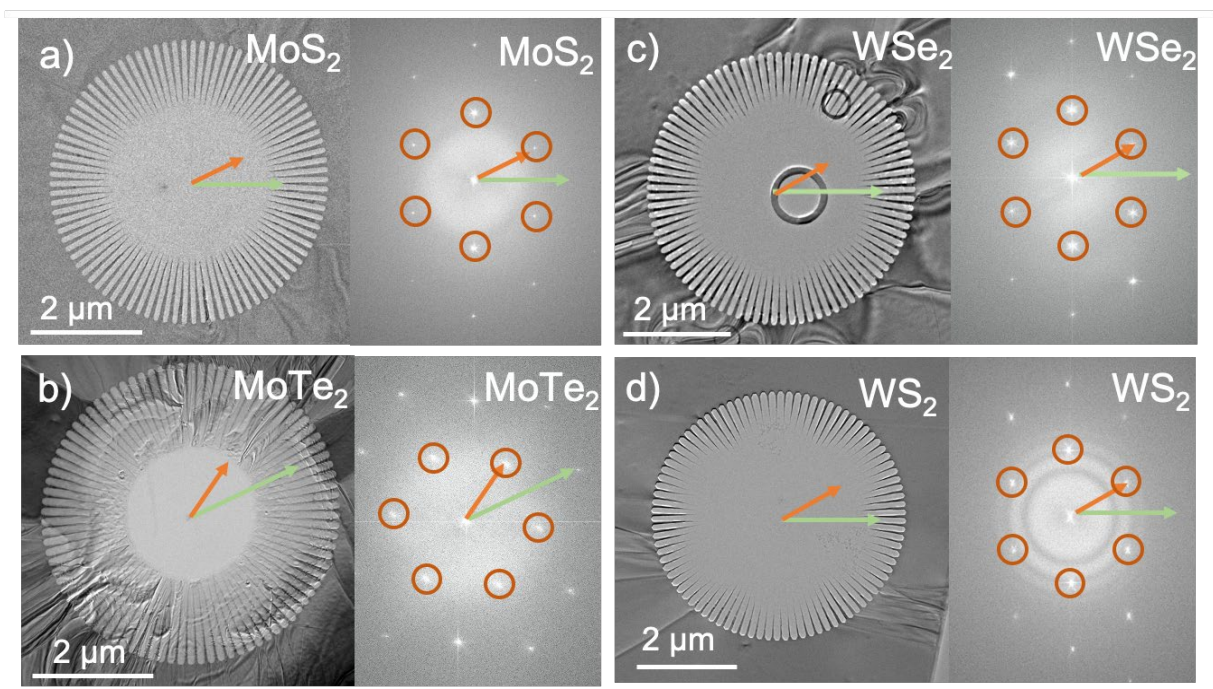

Figure 12. Crystal orientation of wagon wheels, determined from TEM images. (a)-(d) Left, TEM images of WWs in MoS$_2$, MoTe$_2$, WSe$_2$, and WS$_2$, respectively. All WWs have a radius of 2.5 µm. In (c) the rings near the center and the upper right corner are from a bull's-eye structure, but they have migrated to this position during transfer to the TEM grids. (a)-(d) Right, FT TEM images from the same crystal as the WWs to the left. The spots labelled with orange circles belong to the family of 1100 diffraction spots, i.e. they identify the ZZ crystal directions. This means that the ZZ edges are orthogonal to the short orange arrows and the AC edges are orthogonal to the long green arrows.

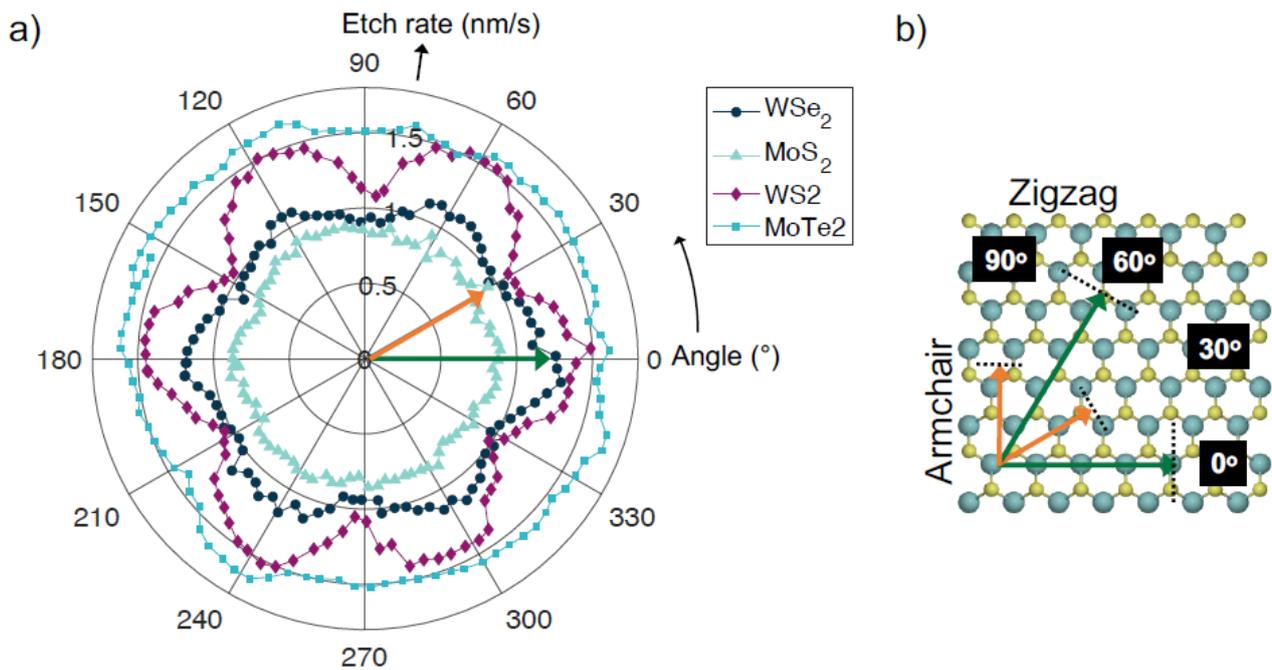

Figure 13 Etch rate vs. crystal orientation. (a)The etch rates as a function of crystal orientation are calculated from TEM images of WWs (see Figure 12). The etch rates for each TMD are calculated from a single wagon wheel. The zero-angle on the graph identifies the crystallographic direction orthogonal to the AC edge for all wagon wheels. (b) Top-view illustration a TMD lattice, with the AC and ZZ directions indicated. This confirms that the fastest etching direction is orthogonal to AC edges (long green arrow), and the slowest etching direction is orthogonal to ZZ edges (short orange arrows).

## 4 Determining Edge Roughness

Supplementary Figure 14a shows the position coordinates of the $WS_2$ edge shown in Supplementary Figure 14b. It is the edge of the inner dot in the bull's-eye structure shown in Figure 3a, thus it is AC-terminated. The Root Mean Square (RMS) roughness based on the raw position coordinates of the edge (blue dots) is 1.3 nm. However, it is clear that the edge is slightly curved, deviating systematically from the true AC facet. We therefore subtract the parabolic deviation from the AC facet to compute an adjusted RMS roughness (see yellow dots and purple mean value). The adjusted RMS roughness, due to random atomic-scale variations is 0.36 nm. The maximum deviation of the edge from the true AC facet (i.e. height of the approximate parabola) is 5 nm.

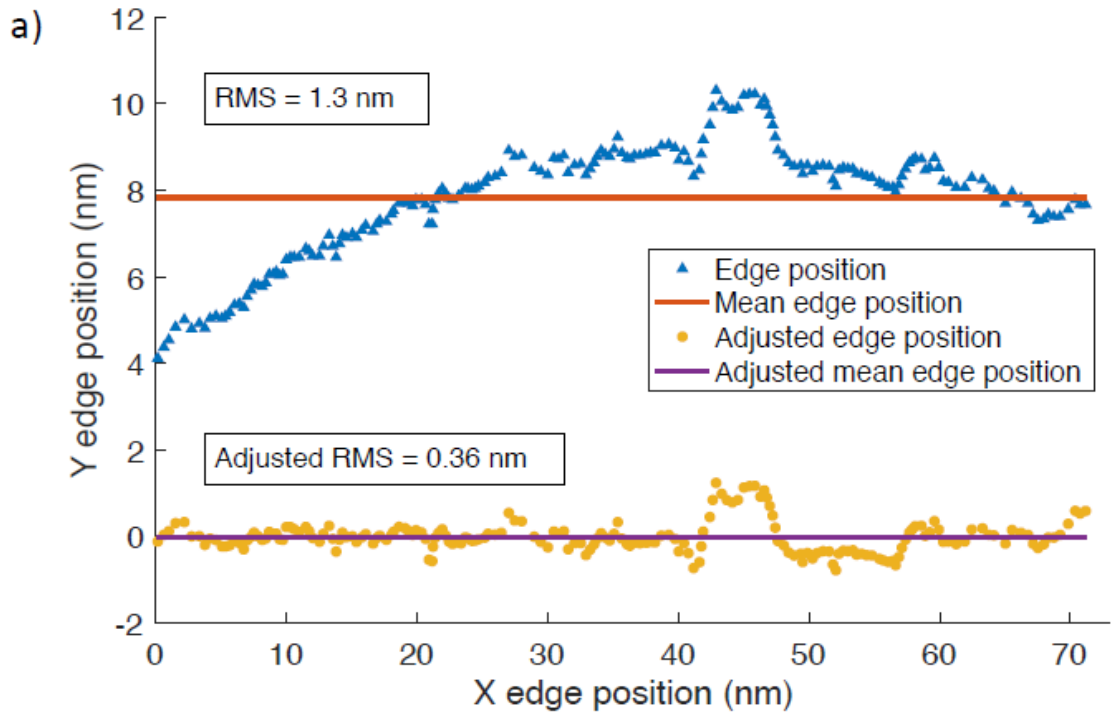

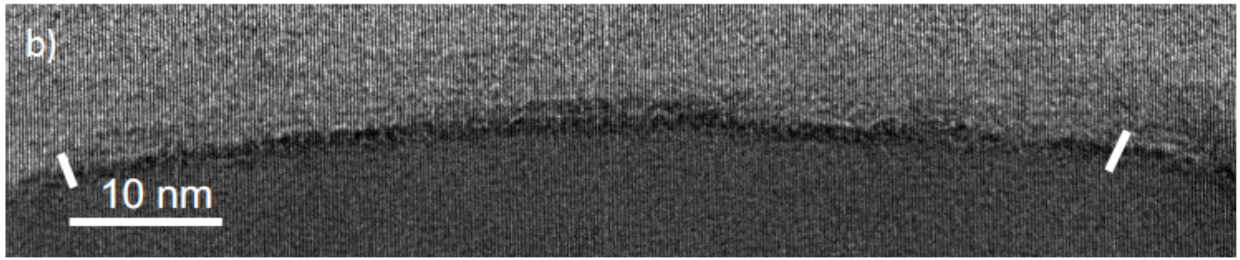

Figure 14. Edge roughness of an AC edge. (a) Position coordinates of the $WS_2$ edge (blue dots), determined from the TEM image in (b). The red line denotes the mean ycoordinate for the edge. The yellow dots is the adjusted edge position, computed by subtracting the systematic curvature of the edge. (b) TEM image of the AC $WS_2$ edge. The edge is from the inner dot in the bull's-eye structure in Figure 3a. The two white lines indicate the considered part of the edge. There are two contributions to the roughness, the parabolic deviation from the true AC facet, and atomic-scale variations of the edge. When determining the adjusted RMS edge roughness, the parabolic deviation from the AC facet is subtracted so only random atomic-scale variations are considered. The RMS roughness based on the raw data of the edge is 1.3nm. The adjusted RMS roughness is 0.36 nm.

For the ZZ edge in Supplementary Figure 20, which is from a hole in the antidot lattice in Supplementary Figure 3f, the RMS roughness is 0.404 nm, the adjusted RMS roughness is 0.225 nm, and the total maximum deviation of the edge from the true ZZ facet is 1.2 nm. The edge roughness might be lower for the ZZ edge because the holes in the $WS_2$ antidot lattice are not etched all the way through the crystal. Previous results[5] show that having an underlying 2D layer reduces edge roughness, when etching hBN with SF$_6$. The smaller deviation from the crystal facet for the ZZ edge, compared to the AC edge, can be explained by difference in feature sizes.

The ZZ holes are much smaller than the AC dot in the bull's-eye, and therefore the ZZ holes are turned into more well-defined hexagons.

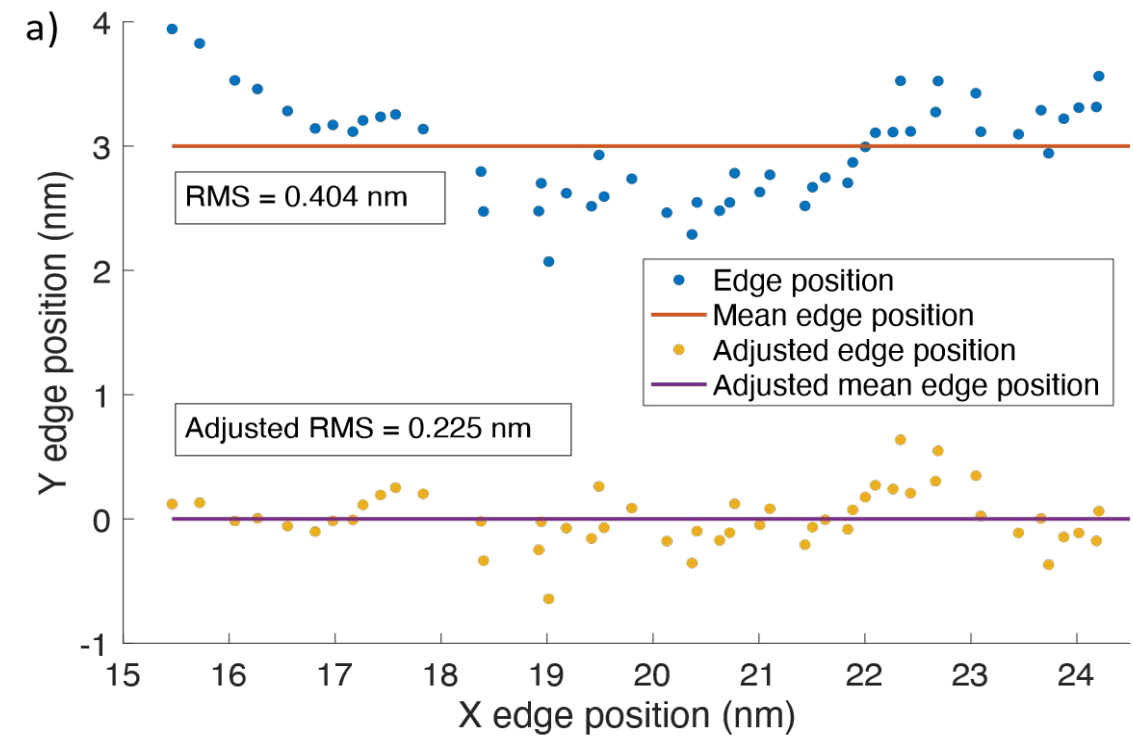

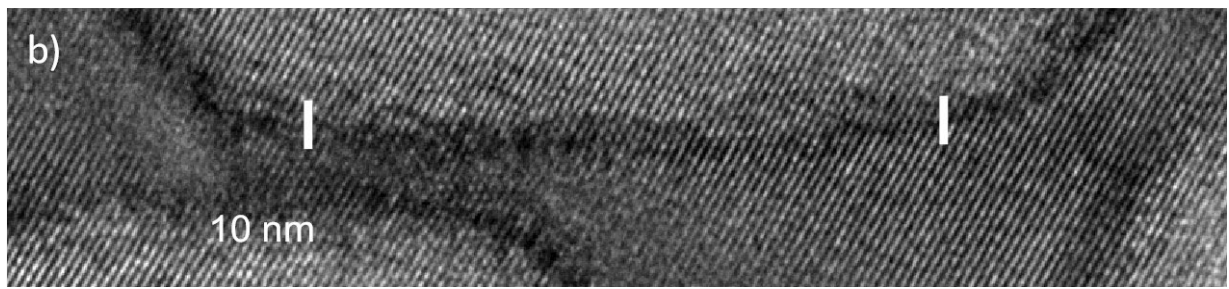

Figure 15. Edge roughness of a ZZ edge. (a) Position coordinates of the WS$_2$ edge (blue dots), determined from the TEM image in (b). The red line denotes the mean y-coordinate for the edge. The RMS edge roughness is 0.404 nm. Yellow dots show the y-coordinates for the edge after adjusting for the curvature, by subtracting the parabolic deviation from the ZZ facet. The purple line denotes the adjusted mean positions of the edge. The adjusted RMS edge roughness is 0.225 nm. The true y position coordinates have been shifted 3 nm upwards in order to separate the two graphs, for clarity. (b) TEM image of the edge. The two white lines indicate the considered part of the edge. The edge is a ZZ edge of a hole in the hexagonal antidot lattice, also shown in Figure 3f.

## 5 Pattern Transfer and Downsizing

Examples of sub 10 nm gaps etched in graphite with downsized hBN hard masks are shown in Supplementary Figure 16 (panel (d) also appears in Main Figure 5). Structures like this could be used for advanced device architectures split gates, photonics, short channel transistors, molecular electronics, etc.

Lithography defined triangles, squares, and hexagons etched into graphite through a 78 nm thick hBN mask are shown in Supplementary Figure 17. The SEM micrographs are obtained after the fabrication steps shown in Supplementary Figure 2f. The side length of the different shapes is noted above the SEM micrographs as a figure of marriage for the dimension.

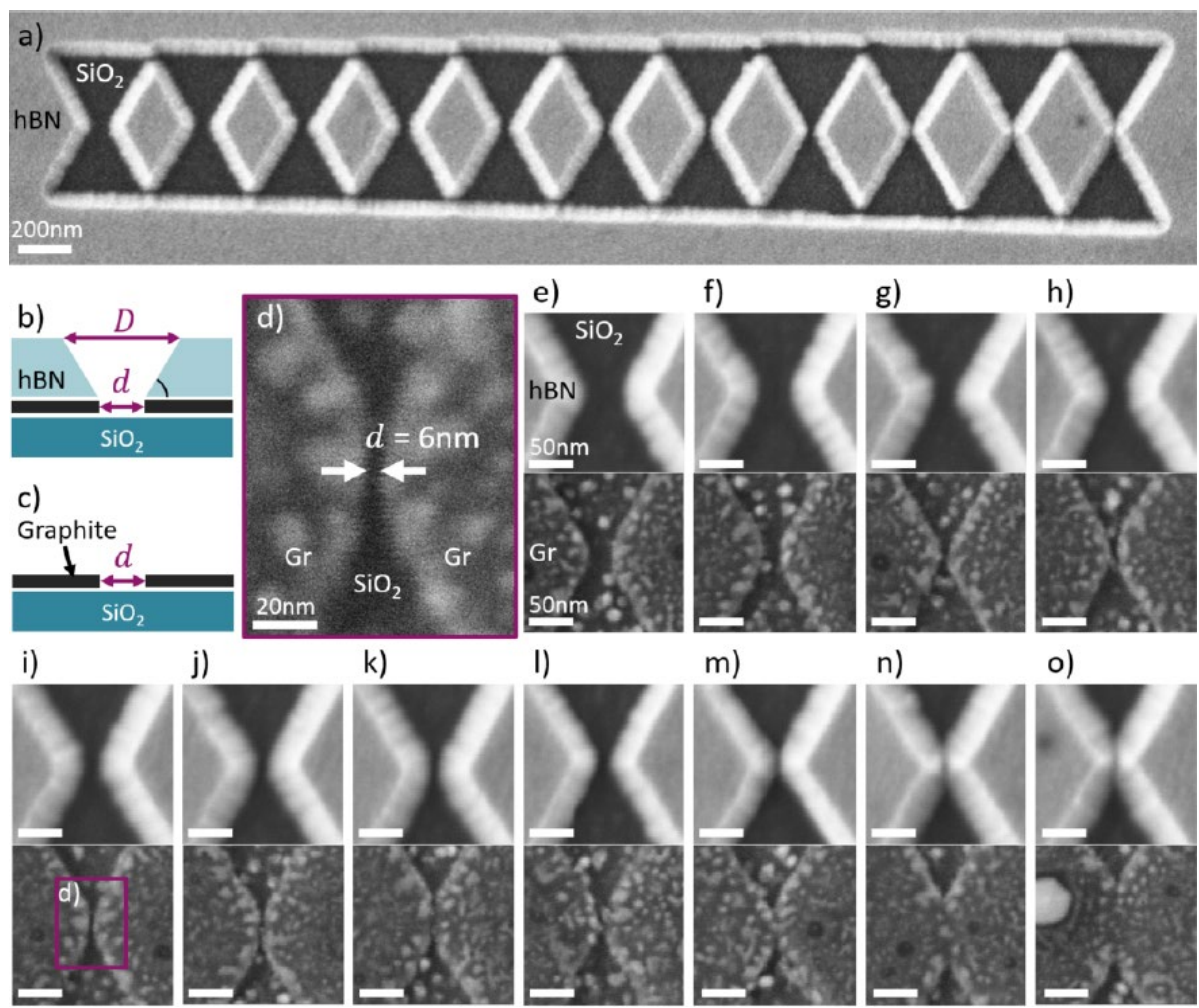

Figure 16. Sub 10 nm gaps in graphite made with an hBN mask. (a) SEM micrographs of a series of test structures with gaps of varying size etched into a stack of graphite-hBN as shown in (b). (e)-(f) A Series of zoom-ins are taken at etch gap both after at the stage illustrated in (b) and (c) where the hBN have been removed with a SF$_6$ etch, some residues are still left on top of the graphite. (d) Gap of 6 nm between two parts of the graphite.

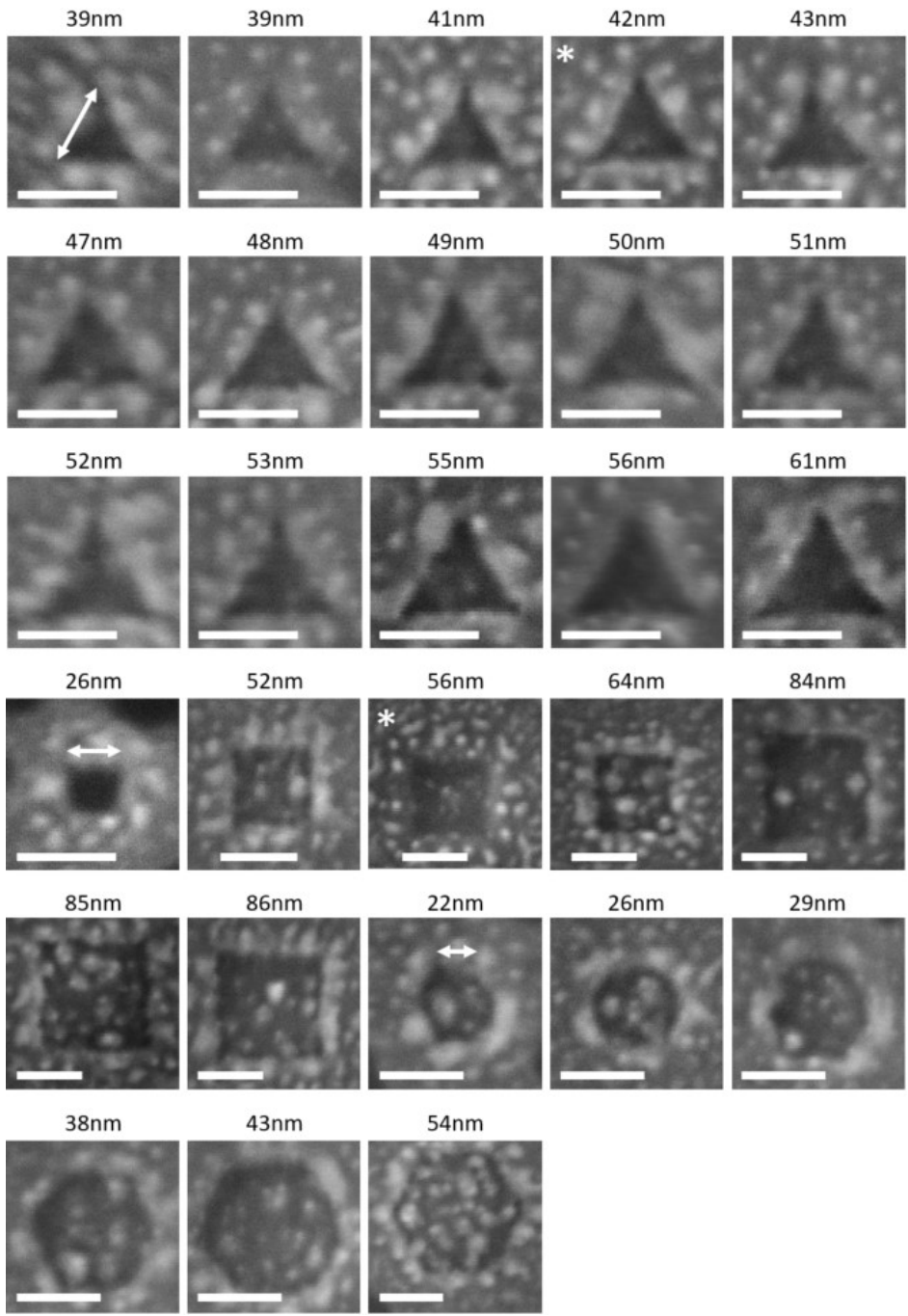

Figure 17. Downsized structures in graphite through hBN. Lithography defined triangles, squares, and hexagons etched in ∼ 2nm thin graphite through a 78 nm thick hBN mask. The dimension marked with the double-ended arrow is noted above the SEM micrographs. Structures marked with "*" are a part of the Main Figure 4. All scale bars are 50 nm.

Supplementary Figure 18 shows SEM micrographs of holes in graphite that have been etched through a layer of WS$_2$ leading to the hole/dot evolving into a hexagonal shape. The WS$_2$ mask have been etched for 90 s with SF$_6$ to ensure the structures are etched fully down to the graphite, the graphite is then etched for 8 min with O$_2$, and the WS$_2$ is finally removed in a 90 s SF$_6$ etch. The SEM inspection is done at the step shown in Supplementary Figure 3f. The etch angle in graphite is in the order of 68° to 77° estimated from the SEM micrographs for the thicker graphite (57 nm) in Supplementary Figure 18. The thickness of the crystals is measured by AFM.

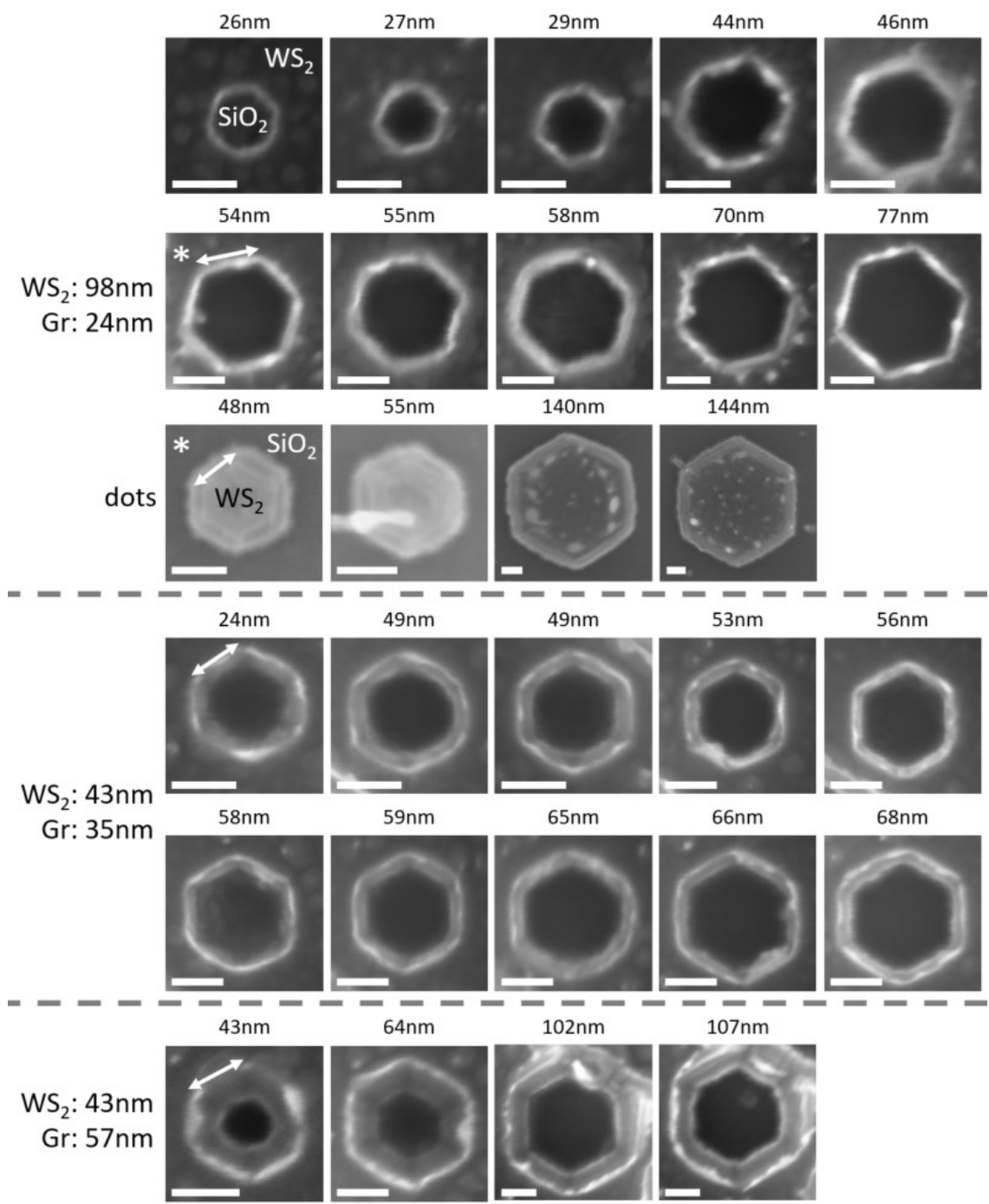

Figure 18. Downsized structures in graphite through $WS_2$. SEM micrographs of hexagonal holes and dots etched into graphite with a $WS_2$ mask, the graphite has three different thicknesses of 24 nm, 35 nm, and 57 nm. The dimension noted above the SEM micrograph is the side length of the hexagon at the top surface of the graphite, indicated by the double-ended arrows. Structures marked with "*" are a part of the main text Figure 4. All scale bars are 50 nm.

# 6 Additional TEM and SEM images

To provide as much information as possible for researchers with interest in confirming, comparing to or further developing our method, we have here added figures that show different characteristic outputs from our etching experiments.

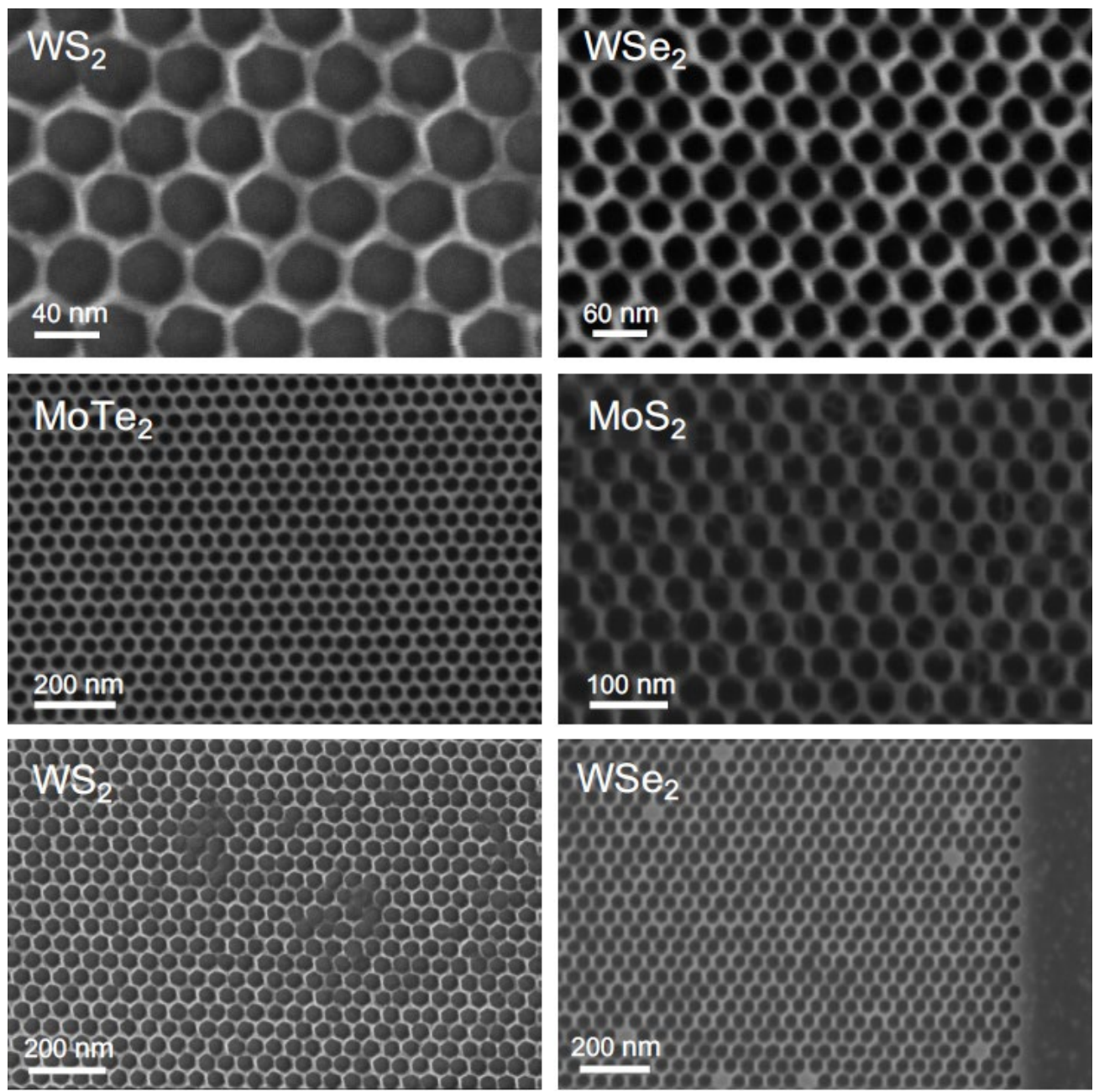

Figure 19. SEM images of antidot lattices in the considered TMD crystals.

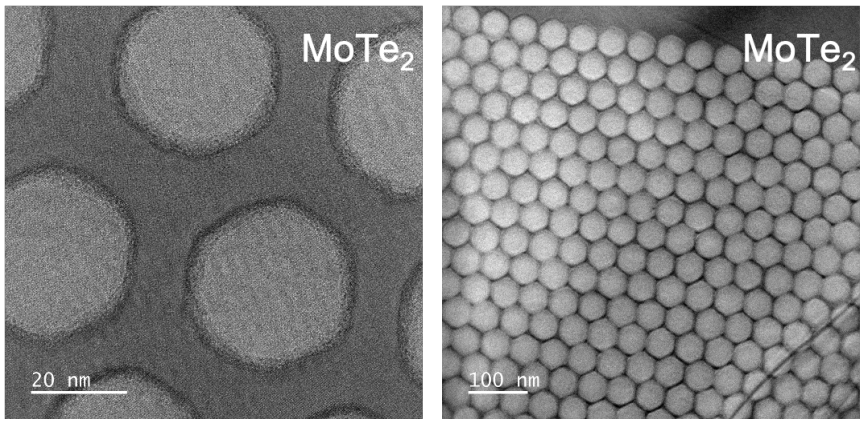

Figure 20: TEM images of an antidot lattice in MoTe$_2$, showing evolving hexagonal ZZ facets

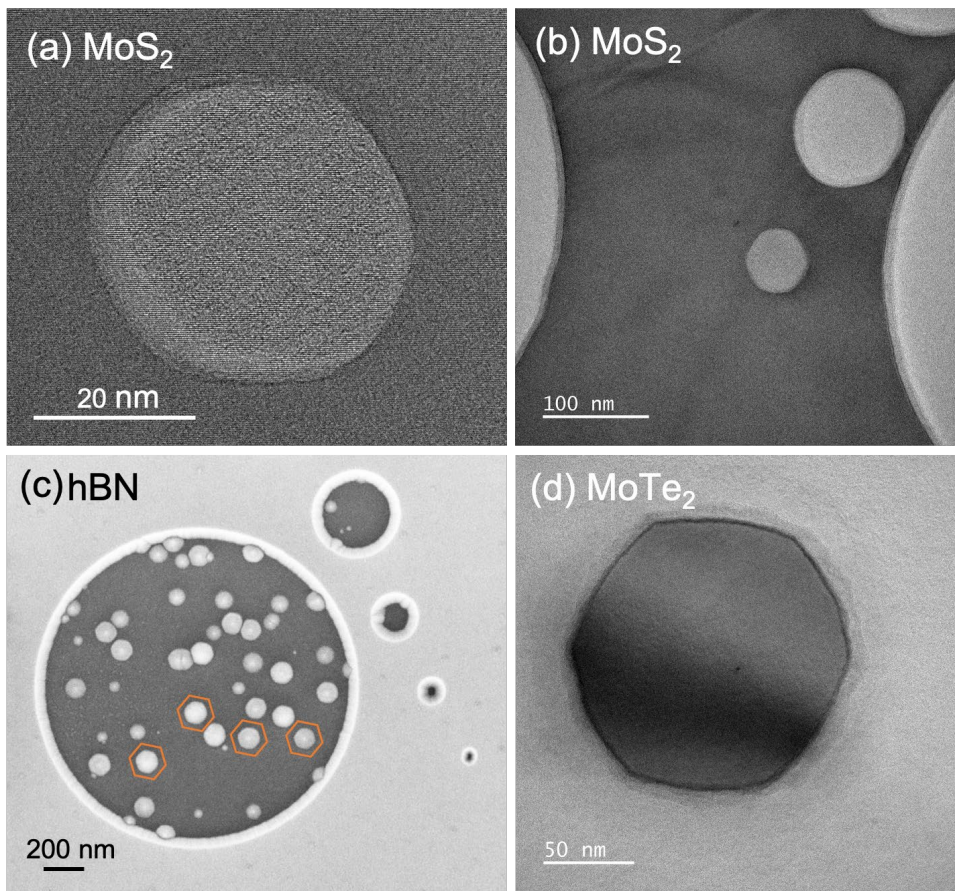

Figure 21. Initial signs of evolving facets in (a)-(b) MoS$_2$ (TEM images), (c) ~ 135 nm thick hBN (SEM image), and (d) MoTe$_2$ (TEM image).

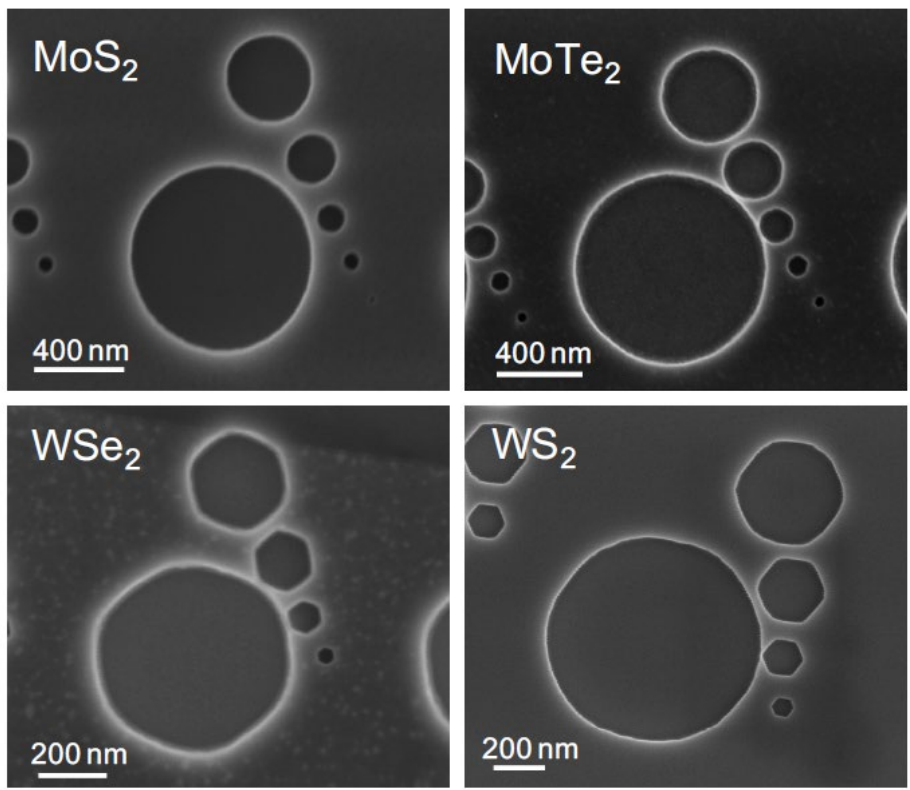

Figure 22. SEM images of etched circles in the considered TMDs. Even the larger circles evolve into hexagons upon etching in WSe$_2$ and WS$_2$. Note the circles have different sizes in the different TMDs.

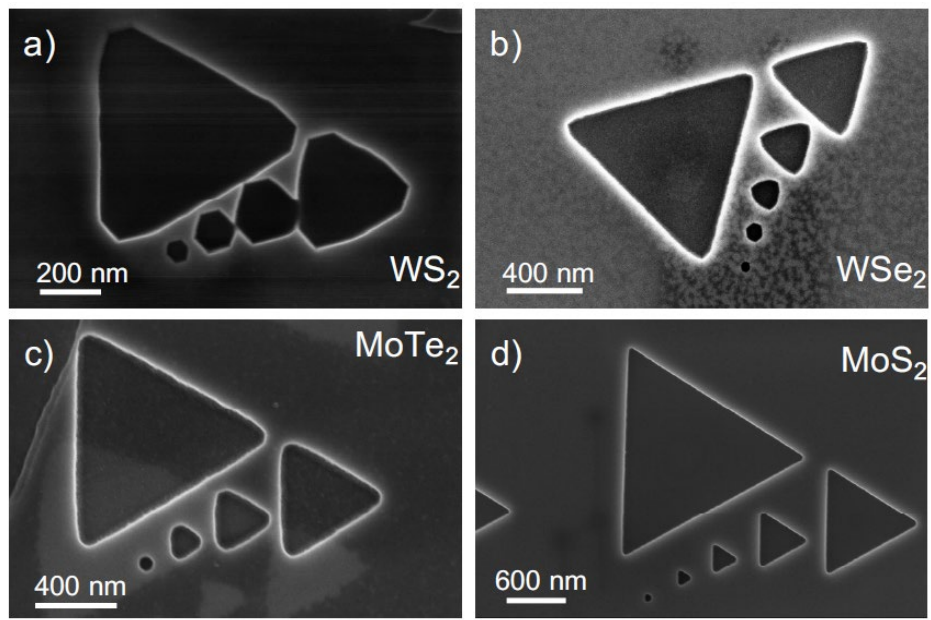

Figure 23 SEM images of etched triangles in the considered TMDs.

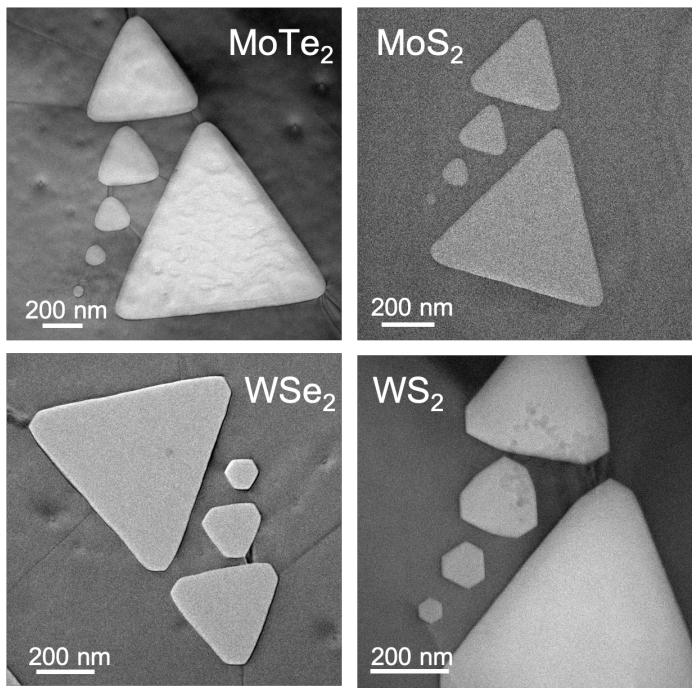

Figure 24. TEM images of etched triangles in the considered TMDs.